\documentclass[reprint, aps, pra]{revtex4-2}
\usepackage{bm, color}
\usepackage{placeins}
\usepackage{nameref}
\usepackage{amssymb}
\usepackage{graphicx}
\usepackage{subfigure}
\usepackage{dcolumn}
\usepackage{bm}

\usepackage{mathtools, nccmath}

\allowdisplaybreaks
\usepackage[colorlinks = true,
            allcolors  = blue]{hyperref}
\usepackage{color,soul}

\usepackage{hyperref}

\preprint{APS/123-QED}
\begin{document}


\title{Predicting nonequilibrium Green's function dynamics and photoemission spectra via nonlinear integral operator learning}

\author{Yuanran Zhu}%
\thanks{Corresponding author: \href{mailto:yzhu4@lbl.gov}{yzhu4@lbl.gov}}
\affiliation{ Applied Mathematics and Computational Research Division, Lawerence Berkeley National Laboratory, Berkeley, USA, 94720.}


\author{Jia Yin}%
\email{jiayin@fudan.edu.cn}
\affiliation{School of Mathematical Sciences, Fudan University, Shanghai, China, 200437.}

\author{Cian C. Reeves}
\email{cianreeves@ucsb.edu}
\affiliation{
Department of Physics, University of California, Santa Barbara,  Santa Barbara, USA, 93117.}

\author{Chao Yang}%
\email{CYang@lbl.gov}
\affiliation{Applied Mathematics and Computational Research Division, Lawerence Berkeley National Laboratory, Berkeley, USA, 94720.}

\author{Vojtěch Vlček}
\email{vlcek@ucsb.edu}
\affiliation{
Department of Chemistry and Biochemistry, University of California, Santa Barbara,  Santa Barbara, USA, 93117.}
\affiliation{Department of Materials, University of California, Santa Barbara,  Santa Barbara, USA, 93117}




\begin{abstract}
Understanding the dynamics of nonequilibrium quantum many-body systems is an important research topic in a wide range of fields across condensed matter physics, quantum optics, and high-energy physics. However, numerical studies of large-scale nonequilibrium phenomena in realistic materials face serious challenges due to intrinsic high-dimensionality of quantum many-body problems and the absence of time-invariance. The nonequilibrium properties of many-body systems can be described by the dynamics of the correlator, or the Green's function of the system, whose time evolution is given by a high-dimensional system of integro-differential equations, known as the Kadanoff-Baym equations (KBEs). The time-convolution term in KBEs, which needs to be recalculated at each time step, makes it difficult to perform long-time numerical simulation. In this paper, we develop an operator-learning framework based on Recurrent Neural Networks (RNNs) to address this challenge.
We utilize RNNs to learn the nonlinear mapping between Green's functions and convolution integrals in KBEs.
By using the learned operators as a surrogate model in the KBE solver, we obtain a general machine-learning scheme for predicting the dynamics of nonequilibrium Green's functions. Besides significant savings per each time step, the new methodology reduces the temporal computational complexity from $O(N_t^3)$ to $O(N_t)$ 
where $N_t$ is the number of steps taken in a simulation, thereby
making it possible to study large many-body problems which are currently infeasible with conventional KBE solvers. 
Through various numerical examples, we demonstrate the effectiveness of the operator-learning based approach in providing accurate predictions of physical observables such as the reduced density matrix and time-resolved photoemission spectra. 
Moreover, our framework exhibits clear numerical convergence and can be easily parallelized, thereby facilitating many possible further developments and applications. 
\end{abstract}

\maketitle
\section{Introduction}
The study of nonequilibrium quantum many-body systems is crucial for understanding a wide range of phenomena in condensed matter physics \citep{krausz2009attosecond,eisert2015quantum,vasseur2016nonequilibrium,golevz2019multiband}, quantum optics \citep{le2016many,giannetti2016ultrafast}, and high-energy physics \citep{kamenev2023field,binder2020dark}. Typical examples include the emergence of transient states phenomena such as quantum phase transitions \citep{dalla2010quantum}, quantum coherence \citep{santos2019role,li2015steady}, and quantum dissipation \citep{breuer2002theory,park2024quasi,huang2024unified}. Despite the importance and urgent need for theoretical study of driven electronic excited states, the intrinsic exponential scaling in the number of degrees of freedom in quantum many-body problems and the lack of time-invariance symmetry in nonequilibrium systems impose serious computational challenges for the investigation of large-scale phenomena in realistic materials. In many-body physics, the prevalent framework for studying electronic excitations is the non-equilibrium Green’s functions (NEGFs)\citep{stefanucci2013nonequilibrium}. Instead of focusing on the many-body wavefunctions, we study the evolution of an effective correlation function that is directly related to experimental observations. For instance, the time resolved photoemission spectroscopy \citep{freericks2008theoretical} directly probes the evolution of individual quasiparticles (QPs), which are electrons and holes dressed by their interactions with the system. The dynamics is given by the (single-particle) Green's function, a two-time correlator $G_{ij}(t,t')$ describing the probability amplitude associated with the QP propagation between sites $i$ and $j$. The time evolution of $G$ is governed by a set of integro-differential equations knowns as the Kadanoff-Baym equations (KBEs)\citep{stefanucci2013nonequilibrium,kadanoff2018quantum}. The KBE is formally exact, but it critically depends on the self-energy, $\Sigma$, representing the downfolded many-body interactions governing the propagation of the QP that is typically unknown. In practice, this effective potential, which is non-local in space and time, is approximated using many-body perturbation theory (MBPT). 

The introduction of Green's function and self-energy can greatly reduce the spatial complexity of the computational problem from the exponential scaling $O(2^{N_s})$ required to represent a wavefunction to $O(N_s^2)$, where $N_s$ is the system size. However, compressing many-body dynamics into an effective single quasiparticle (QP) propagator introduces temporally non-local memory effects, which are represented in the KBEs by a convolution integral of the self-energy. This, so called collision integral,  makes it costly to solve the KBEs numerically.  For nonequilibrium systems, evolving the NEGF $G_{ij}(t,t')$ on a two-time grid requires evaluating the integral at each time step, translating to an asymptotically cubic scaling $O(N_t^3)$ in computational cost, where $N_t$ is the total number of timesteps used in $t$ and $t'$ \citep{reeves2023dynamic,reeves2024real,reeves2023unimportance,blommel2024adaptive,kaye2021low}.  


The temporal non-locality of the collision integral in KBE greatly limits the ability to perform long-time simulations for realistic many-body systems, where $N_s$ is at least $10^3 $ for e.g., driven low-dimensional TMDs \cite{duan2015two}.
In recent years, this limitation has received much attention from both the applied mathematics and condensed matter physics communities. 
New algorithmic approaches to tackle the problem include FFT-based faster solvers \citep{kaye2023fast}, leveraging hierarchical off-diagonal low-rank (HODLR) properties \citep{kaye2021low}, dynamic mode decomposition (DMD) techniques \citep{yin2022using,yin2023analyzing,mejia2023stochastic,maliyov2023dynamic}, and adaptive time stepping strategies \citep{blommel2024adaptive}. These methods focus on developing faster or data-driven algorithms for solving the full KBE. Another prominent class of methods is based on the generalized Kadanoff-Baym ansatz (GKBA), a particular form of time-nonlocal memory truncation, or its reformulations \citep{Lipavsky_1986,tuovinen2020comparing,schlunzen2020achieving,kalvova2024fast,pavlyukh2024cheers,karlsson2021fast,Joost_2020}. These methods could achieve linear time scaling $O(N_t)$ under the GKBA assumption. However, this comes with the trade-off of solving an ODE system of significantly larger dimensionality due to the inclusion of two-electron correlator calculations and the result is less accurate in the time off-diagonal region of $G(t,t')$.

In this work, we propose an RNN-based operator learning framework to address the cubic scaling issue of solving KBEs and develop an effective way to calculate important physical observables such as the reduced density matrix and photoemission spectra. Machine learning (ML) methodologies for learning and predicting complex dynamics have been developed in recent years. Notable techniques include the Physics-informed Neural Network \citep{karniadakis2021physics}, Generative Adversarial Networks \citep{creswell2018generative,wu2020enforcing}, Variational Autoencoders \citep{girin2020dynamical}, and operator-learning frameworks such as DeepONet \citep{lu2021learning} and the Fourier neural operator (FNO) \citep{li2020scalable,kovachki2023neural}. In our previous work \cite{bassi2024learning}, we discovered that  the beside the regular application of RNN in predicting time-dependent dynamics, the architecture can also be used to learn operators, specially the convolution integral operator that has intrinsic memory effects. Hence the general idea we propose here is to use the Long-Short Term Memory (LSTM)-based recurrent neural network \citep{hochreiter1997long} to learn the nonlinear mapping between Green's function $G(t,t')$ and the collision integral $I(t,t') = I[G(t,t')]$ in the KBEs.
After training the neural network (NN) with solutions of the  KBEs obtained in a short time window, we use the RNN as a surrogate model for the collision integral and turn a system of integral differential equations to a system of ordinary differential equations (ODEs) that can be solved efficiently by state-of-the-art ODE solvers with asymptotically $O(N_t)$ computational cost. 
As we shall see, the proposed method yields more accurate prediction of the NEGF dynamics when comparing to data-driven methods such as the DMD. When compared to other linear-scaling solvers such as the GKBA \cite{Lipavsky_1986} and the vectorized solver for non-interacting NEGFs \cite{popescu2016efficient,gaury2014numerical}, our approach avoids the calculation of auxiliary degrees of freedom, such as the two-electron correlator. As a result, it maintains a computational cost comparable to mean-field Hartree-Fock (HF) solvers.

We apply the proposed methodology to various quantum many-body systems and present numerical results that demonstrate the effectiveness of using an RNN-based operator learning framework to predict the dynamics of NEGFs and the time-dependent photoemission spectra of quantum materials.
The NN-based methodology also allows for many potential generalizations and extensions of the framework, thereby offering new possibilities for studying nonequilibrium many-body physics and paving the way for applications in quantum technologies and beyond.




\section{Method}
\subsection{Dynamics of nonequilibrium Green's functions}
We begin by briefly introducing the evolution equation for the NEGFs with technical details provided in Supplementary Note. The Kadanoff-Baym equation (KBE) is a set of integro-differential equations that describe the time evolution of a two-time nonequilibrium Green's function initially at statistical equilibrium and driven by an external field. For a generic many-body system in a lattice:
\begin{equation}\label{MB_ham}
    \mathcal{H}  = \sum_{ij}h_{ij}(t)c^\dagger_ic_j + \frac{1}{2}\sum_{ijkl} w_{ijkl} c^\dagger_ic_j^\dagger c_k c_l,
\end{equation}
where $w_{ijkl}$ is the two-body interaction term and $h_{ij}(t)$ is the single-particle Hamiltonian, the equation of motion (EOM) for KBEs can be formally written as:
\begin{equation}\label{KBE_formal}
\begin{aligned}
\left[i\partial_z - h(z)\right]G(z,z') &= \delta(z,z')  + \int_{\mathcal{C}}\mathrm{d}\bar{z} \Sigma(z,\bar{z})G(\bar{z},z')\\
\left[-i\partial_{z'} - h(z)\right]G(z,z') &= \delta(z,z')  + \int_{\mathcal{C}}\mathrm{d}\bar{z} G(z,\bar{z})\Sigma(\bar{z},z')
\end{aligned}
\end{equation}
Here $G(z,z')=G_{ij}(z,z')$ for $z,z'\in\mathcal{C}$ are NEGFs defined in a contour $\mathcal{C}$ in the complex plane, where $\mathcal{C}$ is known as the Keldysh contour, defined by $\mathcal{C}=\{z\in\mathbb{C}\,|\,\mathrm{Re}[z]\in[0,+\infty],\mathrm{Im}[z]\in[0,-\beta]\}$ \cite{stefanucci2013nonequilibrium}, with $\beta$ being the inverse temperature. $h(z)=h_{ij}(z)$ is the complex-valued single-particle Hamiltonian. The convolution term $I(z,z')=\int_{\mathcal{C}}\mathrm{d}\bar{z} \Sigma(z,\bar{z})G(\bar{z},z')$ is known as the collision integral, where the convolution kernel is the self-energy $\Sigma(t,\bar t)=\Sigma[G](t,\bar t)$ operator which is a nonlinear function of the Green's function according to many-body perturbation theory (MBPT). For moderate and strongly correlated systems where the two-body interaction $w_{ijkl}$ is relatively large, the collision integral has a significant contribution to the Green's function dynamics therefore its accurate approximation and calculation becomes critical for numerical studies of many-body propertities. In this paper, we focus on the parameter range of $w_{ijkl}$ where MBPT remains valid, enabling us to perform full KBE calculations to generate training data. However, the developed methodology is not limited to this regime; it can be readily generalized to predict NEGF dynamics derived from exact diagonalization (ED), where the neural network is trained to learn the {\em exact} collision integral. 

\begin{figure*}
    \centering
    \begin{subfigure}
        \centering
        \includegraphics[width=14cm]{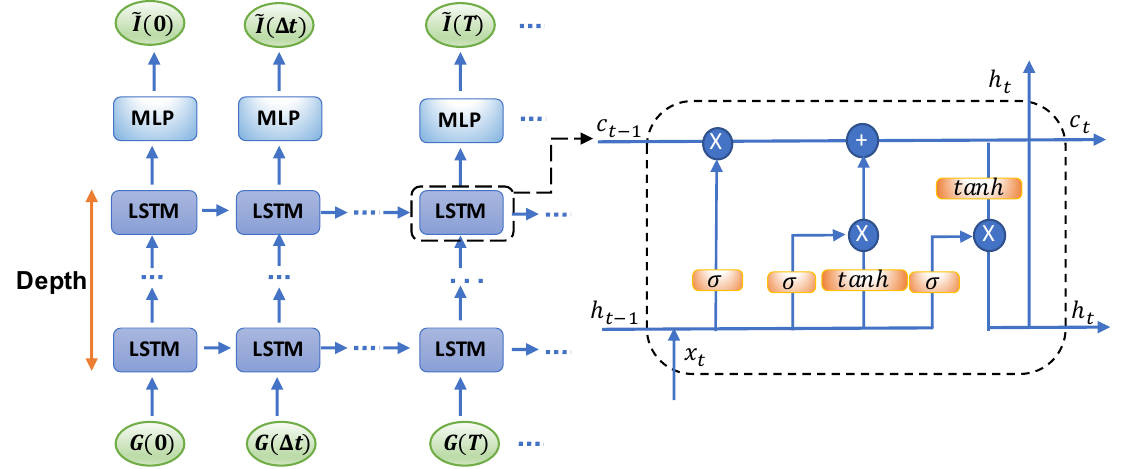}
        \caption{RNN Architecture for learning the nonlinear map between the Green's function and the collision integral: $G\rightarrow I$. The LSTM cells constitute the building blocks of the neural network to capture the memory of the integral operator. The hidden states of the LSTM layers are fed into the last multi-layer perceptron (MLP) layer and then output the approximated collision integral.}
        \label{fig:RNN_architecture}
    \end{subfigure}

        \begin{subfigure}
        \centering
        \includegraphics[width=15cm]{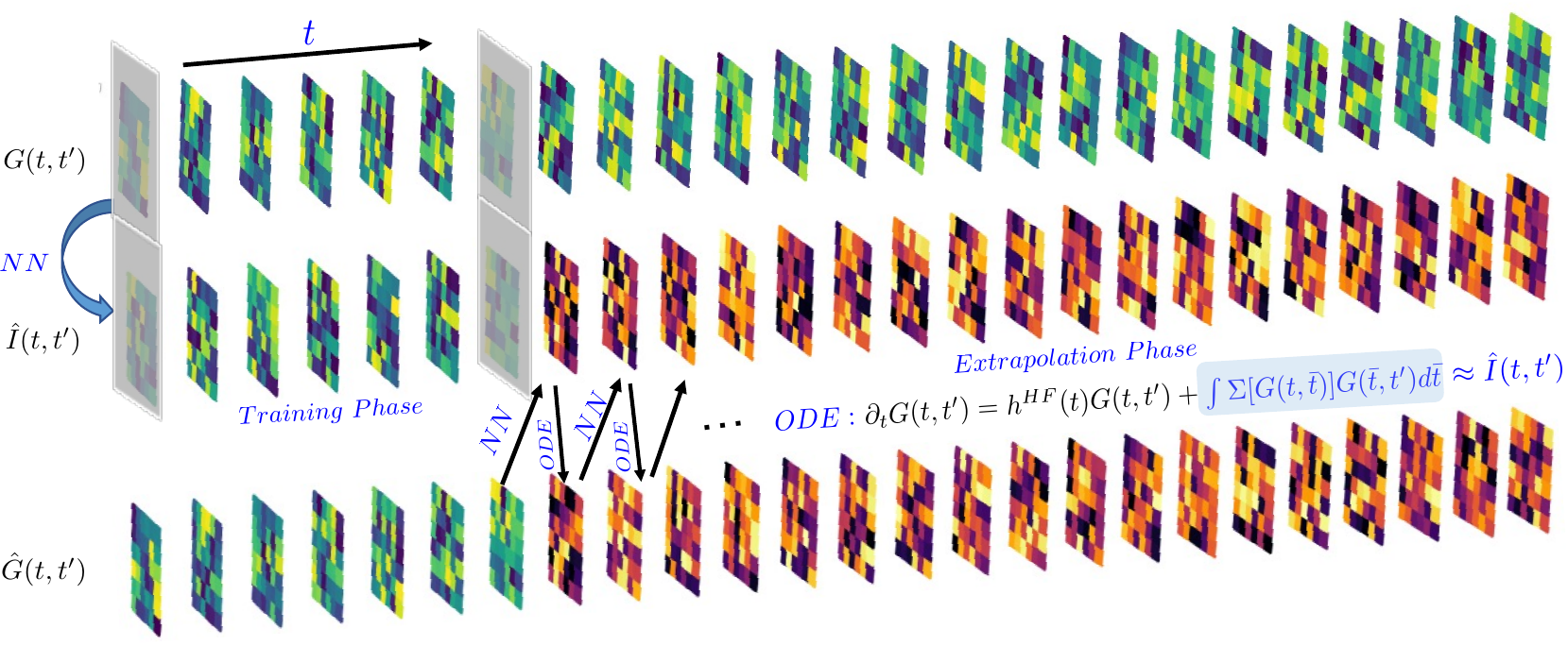}
        \caption{Using the NN-approximated collision integral operator to predict the NEGF dynamics. Here $G(t,t')$ is the ground-truth Green's function obtained by solving KBEs, $\hat I(t,t')$ is the NN predicted collision integral, and $\hat G(t,t')$ is the predicted Green's function using $\hat I(t,t')$. After the training the NN using the KBE solution in a small time window (Training Phase), the NN-predicted collision integral $\hat I(t,t')$ is integrated with the mean-field Hartree-Fock solver \eqref{ODE_extra} to iteratively generate the subsequent timestep Green's function (Extrapolation Phase). Note that throughout the entire procedure, no additional approximations are introduced except the approximation of the collision integral operator $I(t,t')$ by the NN.}
        \label{fig:RNN_extrapolation}
    \end{subfigure}
    \vspace{10pt} 
    \begin{subfigure}
        \centering
        \includegraphics[width=15cm]{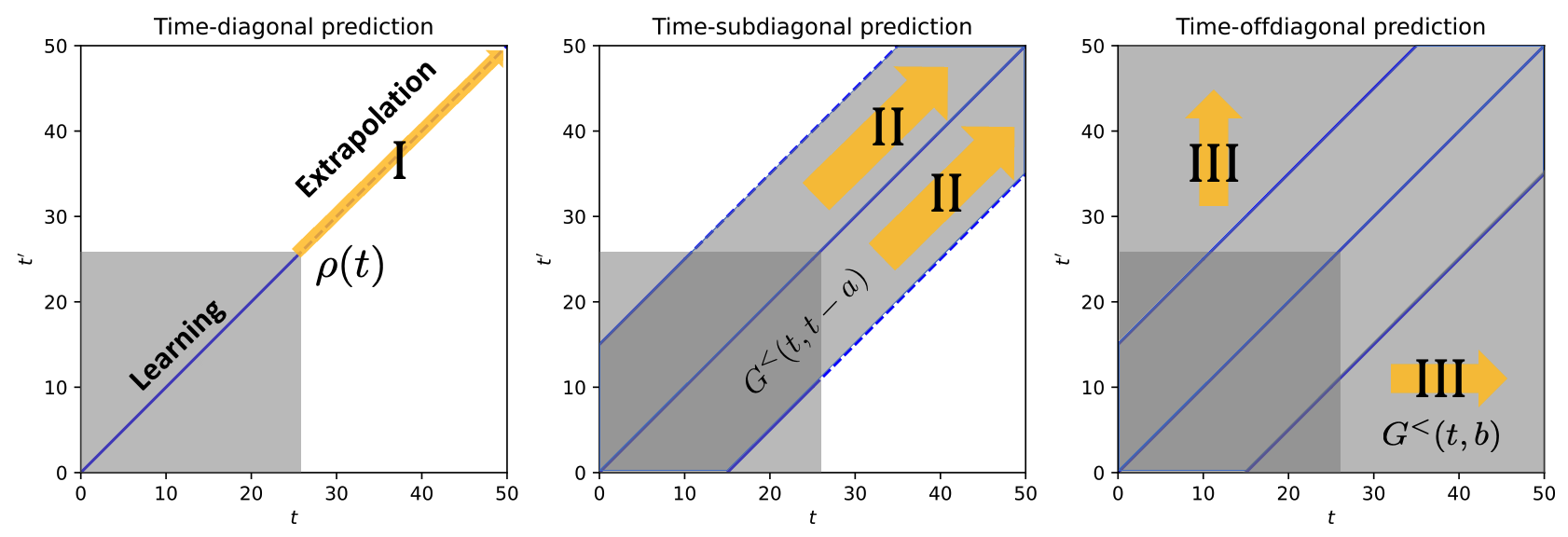}
        \caption{Dynamics reduction of the two-time lesser Green's function into one-time dynamics in the time-diagonal (I), time-subdiagonal (II), and the rest time-offdiagonal regions (III). In each region, we can obtain the reduced EOM (Eqn (3) in Supplementary Note) from the full KBEs for $\rho(t)=-iG^<(t,t)$, $G^<(t,t-a)$, and $G^<(t,b)$ respectively. Note that due to the time-symmetry of the lesser Green's function $G(t,t')=-[G^<(t',t)]^{\dagger}$, we only need to consider the case of $a>0, b>0$ when calculating Green's function in Region II and III.}
        \label{fig:dynamics_reduction}
    \end{subfigure}
\end{figure*}

The complex-valued KBEs \eqref{KBE_formal} can be reformulated into a system of integro-differential equations by introducing Green's functions and self-energies evaluated at different branches of the Keldysh contour. The detailed equation is given in the Supplementary note. For subsequent discussion, it is sufficient to only focus on the EOM for the {\em lesser} Green's function $G^{<}(t,t')$, which is given by
\footnote{Due to the symmetry of the lesser Green's function, in practice, we only need to solve one of the Eqn \eqref{KBE} for off-diagonal dynamics calculations, as explained in the caption of FIG \ref{fig:dynamics_reduction}}
:
\begin{equation}\label{KBE}
           \begin{split}
                i\partial_t G^{<}(t,t') &= h^{\textrm{HF}}(t)G^{<}(t,t') + I_1^{<}(t,t')\\
                -i\partial_{t'} G^{<}(t,t') &= G^{<}(t,t')h^{\textrm{HF}}(t') + I_2^{<}(t,t').\\
            \end{split}
        \end{equation}
Here $G^{<}(t,t') = G_{ij}^{<}(t,t')$ for $t,t'\in \mathbb{R}$ is a four-rank tensors with dimensionality $N_s\times N_s\times N_t\times N_t$, where $N_s$ is the lattice size and $N_t$ is the total number of timesteps. $h^{\textrm{HF}}(t)$ is the mean-field Hamiltonian, where $\textrm{HF}$ stands for the Hartree-Fock approximation. $I^<_{1,2}(t,t')$ are the corresponding collision integrals.

The lesser Green's function $G_{ij}^{<}(t,t')$ encodes important dynamical information of the nonequilibrium system. In this work, we shall focus on two physical observables which are related to the time-diagonal and off-diagonal components of Green's function. The first one is the reduced density matrix defined by
\begin{align*}
\rho(t) = -iG^<(t,t),
\end{align*}
which provides the averaged electron occupation number and the coherence information of the system. Another one is time-resolved spectral function \cite{freericks2008theoretical,reeves2024real}:
\begin{align}\label{noneq_spectral_function}
A(\omega, k, t_p)=\int dtdt' e^{i\omega(t-t')}S_{\sigma}(t-t_p)S_{\sigma}(t'-t_p)
G^<_k(t,t'),
\end{align}
where $G^<_k(t,t')$ is the momentum space lesser Green's function, which is obtained from $G_{ij}^<(t,t')$ via discrete Fourier transform \cite{rusakov2016self}, and the window function $S_{\sigma}(t-t_p)$ determines the energy and temporal resolution of the non-equilibrium spectral function. $A(\omega, k, t_p)$ corresponds to the measured photoemission spectrum in time-resolved photoemission spectroscopy (TR-PES) experiments and describes the non-equilibrium distribution of quasiparticles in energy and momentum space around some probe time $t_p$ after the system is driven out of equilibrium. In this paper, we choose the probing window $S_{\sigma}(t-t_p)$ to be a Gaussian shape function concentrating on the dynamics probing time $t_p$ \cite{freericks2008theoretical}:
\begin{align*}
S_{\sigma}(t-t_p) = \frac{1}{\sigma \sqrt{2\pi}} e^{-\frac{(t-t_p)^2}{2\sigma^2}}.
\end{align*}
From this definition, we see that the Green's function components that are close to the time-diagonals contribute more to the spectral function dynamics. This fact will be used in later computation. 

    
    

\subsection{Operator learning and dynamics reduction}
The operator-learning approach is devised to effective solve the 
KBE \eqref{KBE} for lesser Green's function and calculate the reduced density matrix $\rho(t)$ and time-dependent spectral function $A(\omega, k, t_p)$. The workflow is divided into two parts: learning and predicting KBE dynamics, and dynamics reduction of KBEs, which are explained separately in FIG \ref{fig:RNN_architecture}-\ref{fig:RNN_extrapolation} and FIG \ref{fig:dynamics_reduction}. In implementations, we actually {\em first} apply the dynamics reduction and then use the NN to predict the NEGF dynamics. However, for better explain the motivation behind the dynamics reduction procedure, let us first focus on the learning/predicting portion of the workflow. 

The basic rationale behind our construction is that the main computational cost for solving KBE comes from the evaluation of the collision integral. If $I(t,t')$ can be approximated by a surrogate model without evaluating the self-energy and performing the numerical integration, then solving the KBEs is as cheap as solving an ODE. It follows from renormalized MBPT that $I(t,t')=\int_0^td\bar t\Sigma[G](t,\bar t)G(\bar t,t')$ is a nonlinear integral operator that maps $G(t,t')\rightarrow I(t,t')$. The time-convolution structure motivates us to use LSTM-based RNN to learn the mapping between $G$ and $I$ (FIG \ref{fig:RNN_architecture}). After training the RNN using the KBE solution within a short time window, we use the RNN-approximated mapping to predict $I(t,t')$ and eventually solve for Green's functions (FIG \ref{fig:RNN_extrapolation}). Using the simplest Euler-forward scheme as an example, This is done by solving the following EOM:
\begin{equation}\label{ODE_extra}
\begin{aligned}
iG^{<}(t+\Delta t,t') &=  G^{<}(t,t')\\
&+\Delta t[h^{\textrm{HF}}(t)G^{<}(t,t')+\hat I^{<}_1(t,t')]\\
-iG^{<}(t,t'+\Delta t') &=  G^{<}(t,t')\\
&+\Delta t'[G^{<}(t,t')h^{\textrm{HF}}(t')+\hat I^{<}_2(t,t')],
\end{aligned}
\end{equation}
where $\hat I_{1,2}^{<}(t,t')$ is the output of the RNN obtained by taking $G^{<}(t,t')$ as the input. 

What distinguishes the approach taken in this work and other previous approaches for learning the NEGF dynamics is that we do not learn the dynamics of NEGF directly. 
There are two benefits for doing this. First, the mapping between $G(t,t')$ and $I(t,t')$ is {\em univeral} for different NEGF inputs, i.e., the operator itself is independent of the function input $G(t,t')$
\footnote{As an example, consider Hubbard model with different $U$ values. The 2nd-order Born self-energy has always been $U^2G(t,t')\circ G(t,t')\circ G^T(t',t)$, where $\circ$ denotes the Hadamard product, hence of the corresponding collision integral.} .
As a result, once such a mapping is well approximated, we can use the learned operator to solve for NEGFs corresponding to systems driven by different one-body forces. Secondly, it is also numerically beneficial to make predictions of $I(t,t')$ because for systems with weak or moderate interaction strength, the magnitude of $I(t,t')$ is always much smaller than that of $G(t,t')$ \cite{reeves2023unimportance}. This means that with roughly the same order of relative approximation accuracy, using the predicted $I(t,t')$ to solve for $G(t,t')$ will lead to more accurate result than learning, say $\partial_t G(t,t')$. 


We now turn to the dynamics redution procedure. When implementing the RNN, we found that training a NN that takes a four-rank tensor $G^{<}_{ij}(t,t')$ as the input and output another one i.e. $\hat I^{<}_{ij}(t,t')$, is computationally costly and also unnecessary for calculating physical quantities like the photoemission spectra $A(\omega,k,t_p)$ (explained below). Therefore, the second part of the  workflow consists of a dynamics reduction procedure \cite{yin2022using,yin2023analyzing} for KBEs which enables learning and predicting the one-time dynamics of NEGFs. Specifically, from the full KBEs, we can get the reduced EOM for different one-time Green's functions: $G^<(t,t), G^<(t,t-a)$ and $G^<(t,b)$ (see FIG \ref{fig:dynamics_reduction} for schematic illustrations and Supplementary Note Section 1 for the specific form of the reduced EOMs). This procedure divides the learning/predicting tasks into three parts which can be executed one-by-one. Specifically, for the first step, we learn and solve for Green's function along the time-diagonal to get $G^<(t,t)$ (Region I in FIG \ref{fig:dynamics_reduction}). Then we repeat the procedure for the time-subdiagonal dynamics to get $G^<(t,t-a)$ for different $a$ (Region II in FIG \ref{fig:dynamics_reduction}). Lastly, we perform the computation along other time off-diagonals to get $G^<(t,b)$ for different $b$ (Region III in FIG \ref{fig:dynamics_reduction}). The dynamics prediction in Region III is less accurate due to the long-term memory effects in the off-diagonal regime (c.f. Supplementary Note Section 2.3). 
However, for calculating $A(\omega,k,t_p)$ whose main contribution comes from the time-subdiagonal part (Region II) of Green's function $G^<(t,t')$ according to its definition \eqref{noneq_spectral_function}, the calculation in Region III can therefore be omitted and we still get an accurate prediction of the photoemission spectra (see FIG \ref{fig:Extrapolation_diagram_heatmap}-\ref{fig:band_structure}).
Further explanations of the dynamics reduction procedure, choice of time integrators, and the reasoning behind all the construction can be found in Section \ref{sec:Methods}.
\subsection{Results}\label{sec:Results}
We test the proposed operator-learning method on predicting NEGF dynamics of interactive systems that have various sites, interaction strengths, and perturbations induced by different nonequilibrium forces. To benchmark our result, we compare the extrapolated lesser Green's function $G^<(t,t')$ with the ground-truth full KBE solution. In the supplementary note, we further provide systematical comparative studies with approximated KBE solutions generated by two other approaches: the time-dependent Hartree Fock (TDHF) and the dynamic mode decomposition (DMD).



%
%
%
%
\begin{figure*}
\centering
\begin{subfigure}
\centering
\includegraphics[width=15cm]{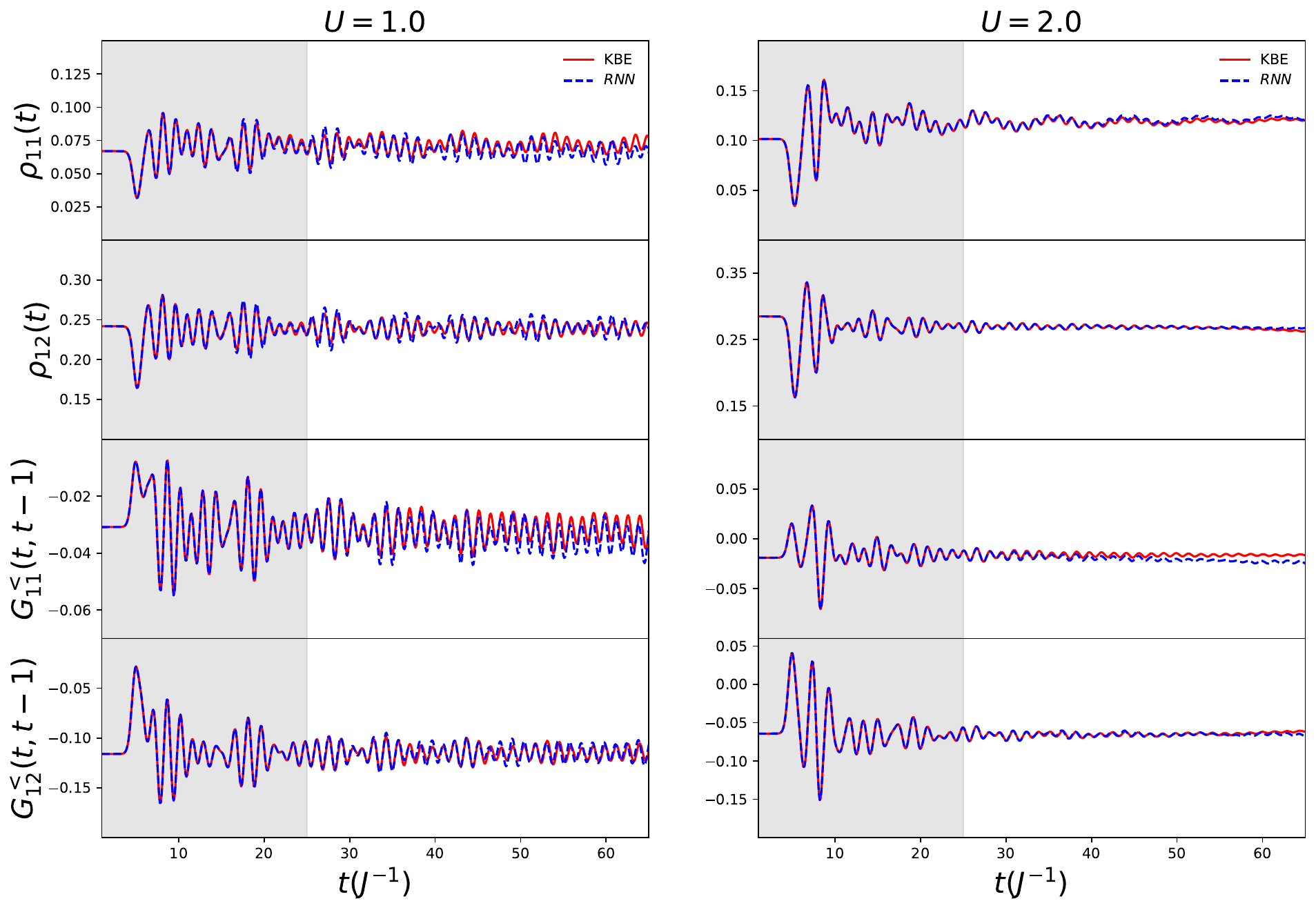}
\end{subfigure}
\caption{Predicted reduced density matrix $\rho(t)=-iG^<(t,t)$ and time-subdiagonal lesser Green's function $G^<(t,t-1)$ generated by RNN compared with the ground truth KBE solution. Here we only plot the imaginary part of the matrix element. The displayed result is for Hubbard model with $U=1.0J$ and $U=2.0J$.}
\label{fig:N4_extrapolation}
\end{figure*}
\subsubsection{Real-time dynamics}
\begin{figure*}
\centering
\includegraphics[width=17cm]{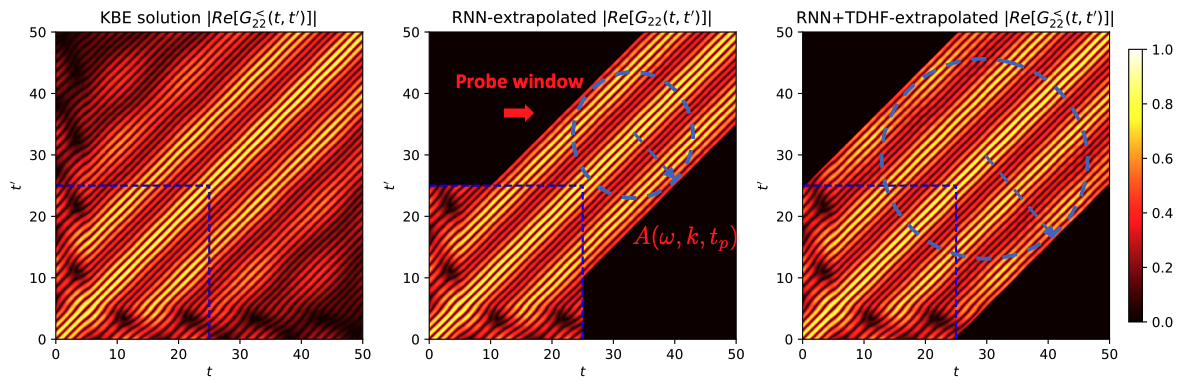}
\caption{Predicted one-time Green's function dynamics in different dynamics reduction regions. The obtained time-subdiagonal data can be used to calculate the nonequilibrium spectral function $A(\omega,k,t_p)$. If a larger probing window is needed for calculating $A(\omega,k,t_p)$ with higher temporal resolution, one may use the TDHF data to fill the dynamics in the region indicated by the third figure (Details in Section \ref{sec:extrapolation}).}
\label{fig:Extrapolation_diagram_heatmap}
\end{figure*}

We first compare the RNN-predicted real-time dynamics $G^<(t,t')$ with the KBE result. Specifically, we choose a Hubbard-type interaction $w_{ijkl}=U\delta_{ijkl}\delta_{i\uparrow,i\downarrow}$ in the modeling Hamiltonian \eqref{MB_ham}. The nonequilibrium forces are added in the single-particle Hamiltonian $h_{ij}(t)$ \cite{reeves2023dynamic,reeves2023unimportance,reeves2024real}:
\begin{align*}
h_{ij}(t) = h^{(0)}_{ij}+h_{ij}^{N.E}(t),
\end{align*}
where 
\begin{align*}
h^{(0)}_{ij} = J(\delta_{i,j-1}+\delta_{i,j+1}) +V(-1)^i\delta_{ij},
\end{align*}
and the nonequilibrium driving term is 
\begin{align*}
h_{ij}^{N.E}(t)=
\begin{cases}
\delta_{ij}E\cos(\pi r_i)\exp\{-\frac{(t-t_0)^2}{2T_p^2}\}\\ 
\delta_{ij}Er_i\exp\{-\frac{(t-t_0)^2}{2T_p^2}\}
\end{cases}
\end{align*}
where $r_i=\frac{1}{2}\left(\frac{N_s-1}{2}-i\right)$, $N_s$ is the lattice size. We will call the first case {\em short wavelength} potential force due to the existence of the $\cos(\pi r_i)$ function that modulates the pulse on the lengthscale of individual sites and the second kind {\em long wavelength} potential force. The modeling parameters $\{J,U,V,N_s,E,T_p\}$ will be determined later for different simulations. In FIG \ref{fig:N4_extrapolation}, we show the dynamics prediction result for a 4-site Hubbard model driven by the long wavelength force with modeling parameters $\{J=1,U=1/2,V=2,N_s=4,E=1.0, T_p= 0.5\}$ at the inverse temperature $\beta =20$. This relatively small system is used merely for illustration purposes and the methodology is readily generalizable to much larger systems as we will show in the next section.
To train the RNN, we use the KBE solution in the time-domain $t\in [0,25](J^{-1})$ as the training data, and this is indicated by the shaded region in subfigures. From FIG \ref{fig:N4_extrapolation}, we see that the operator-learning approach yields accurate predictions of the Green's function dynamics in both the time-diagonal and time-subdigonal region. We also gathered the extrapolation results of $G^<(t,t-a)$ for different $a$ and show them in FIG \ref{fig:Extrapolation_diagram_heatmap}. 

We further tested our approach for Hubbard model with different sites (e.g. $N_s=8,12$) and driven by different external fields. All these test results, as well as the comparison with the TDHF and DMD approach are included in Supplementary Note Section 2. The general conclusion we obtained by the comparative study is that the operator-learning method consistently yield the most accurate and reliable predictions of Green's function dynamics, both in the time-diagonal and time-subdiagonal regions. In particular, when comparing with the TDHF result (c.f. Figure 1-4 in Supplementary Note Section 2), the dynamics prediction gets significantly improved for Hubbard model with larger interaction strength $U=2.0$, which clearly indicate that the RNN well-captured the memory effect caused by the many-body correlation. When comparing with DMD approach, the operator-learning method requires less training data and makes more accurate prediction of the Green's function dynamics. Moreover, as an operator, the collision integral operator mapping $G\rightarrow I$ is {\em universal} for different Green's function inputs once the self-energy approximation is chosen. Hence the operator learned for one system can be transferred to predict the dynamics of the Hubbard model driven by different one-body forces. 

In addition to the quantitatively assessment for the quality of dynamics prediction, we also provide numerical convergence analysis for the proposed methodology in Supplementary Note. As a data-driven method, it is typically hard to predetermine how much data is enough for constructing a reliable prediction of dynamics when the future dynamics in principle are unknown. To address this issue in the operator-learning framework, we performed a series of numerical convergence tests for different systems and demonstrated that the operator-learning approach yields consistent dynamics predictions that numerically converge to the true solution. This leads to a straightforward and computationally efficient adaptive procedure to predetermine the required training data for dynamics extrapolations (c.f. Supplementary Note Section 3). Gathering all the test results, we claim that the RNN indeed provide reliable predictions of the {\em relaxation} Green's function dynamics after the imposing of the quench force. 
\subsubsection{Photoemission spectrum}
\label{sec:emission_spectra}
%
%

\begin{figure*}
\centering
\includegraphics[width=17cm]{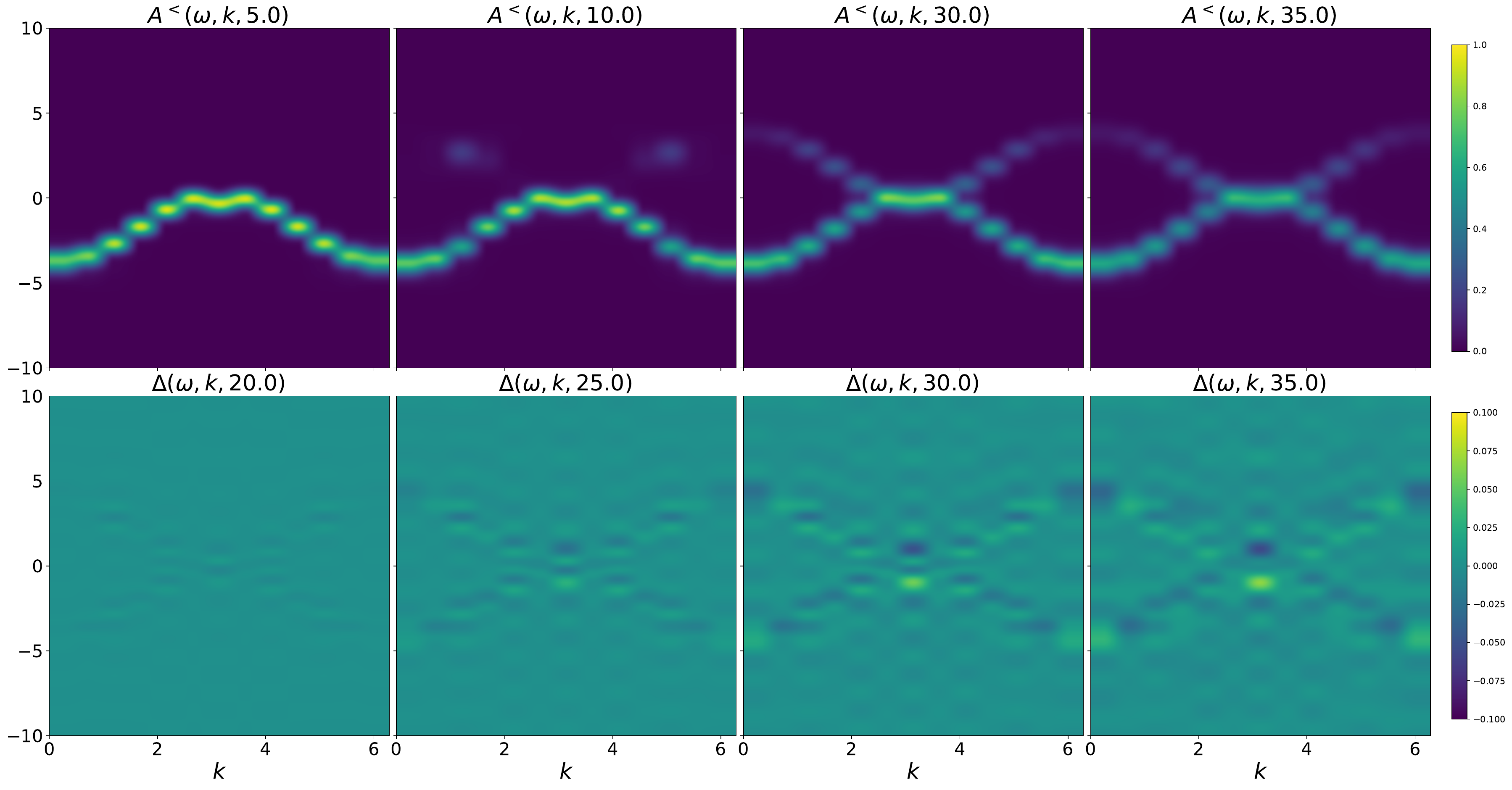}
\caption{Time evolution of the nonequilibrium spectral function $A^<(\omega, k, t_p)$ for the two-band Hubbard model predicted by
the RNN (first row). The training KBE data is collected for $0\leq t,t'\leq 20$. In the second row, we display the dynamics prediction error $\Delta(\omega,k,t_p) =A_{KBE}^<(\omega,k,t_p)-A_{RNN}^<(\omega,k,t_p)$ of the RNN approach for different probing time $t_p$.}
\label{fig:band_structure}
\end{figure*}

In this section, we show how the operator-learning approach works in predicting the time evolution of the photoemission spectra $A(\omega, k, t_p)$ -- and observable containing information on both time diagonal and off-diagonal elements of Green's function. Specifically, we consider the driven dynamics of the semimetal modeled by the full Hamiltonian:
\begin{align*}
\mathcal{H} &= \sum_{\alpha,\beta\in\{c,v\}}\sum_{\langle ij\rangle\sigma}h_{ij}^{\alpha,\beta}(t)c_{i\sigma}^{\alpha\dagger}
c_{j\sigma}^{\beta}-\sum_{\alpha\in\{c,v\}}\mu_{\alpha}n^{\alpha}_{i\sigma}\\
&\ \ \ +U\sum_{i,\alpha\in\{c,v\}}n^{\alpha}_{i\uparrow}n^{\alpha}_{i\downarrow}\\
h_{ij}^{\alpha\beta}(t) &= J\delta_{\alpha\beta}\delta_{\langle i,j\rangle}\\
&\ \ \ +\delta_{ij}(1-\delta_{\alpha\beta})E\cos(\omega_p(t-t_0))e^{-\frac{(t-t_0)^2}{2T_p^2}},
\end{align*}
which corresponds to a Hubbard model with two orbitals per site and an orbital energy $\mu_\alpha$ for $\alpha \in \{c,v\}$, with $\{c,v\}$ representing the conduction/valence band. In the Hamiltonian, the first term describes the kinetic energy and the time-dependent driving, where $\langle i,j\rangle$ implies nearest neighborhood hopping and the perturbation is an optical pulse that pumps electrons from the valence to the conduction band. The third term describes an onsite interaction between opposite spin particles, which is characterized by the parameter $U$. Finally, we impose periodic boundary conditions to the system. Other modeling parameters are chosen to be: $\{N_s= 12, J =1.0, U=1.0, \mu_v=-2.0,\mu_c=2.0, \omega_p=5.0, t_0 = 10.0, T_p= 1.0, \beta=20\}$. The probe width $\sigma$ of the window function $S_{\sigma}(t-t_p)$ is specified to be $\sigma =2$. The RNN-predicted spectral function is shown in FIG \ref{fig:band_structure}. From the figure, we see that it provides quantitatively accurate prediction of the dynamical process of how the added perturbation force creates electrons in the conduction band and then drive the excitation to populate the whole band. 
%
%
\subsubsection{Computational cost and scalability}
\label{sec:computational_cost}
\begin{figure}
\centering
\includegraphics[width=8cm]{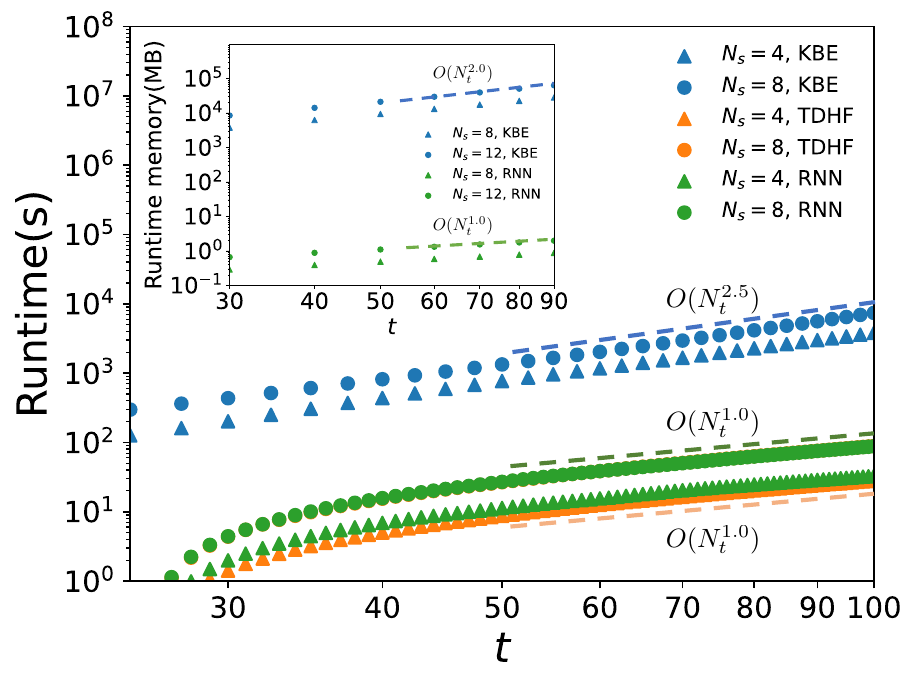}
\caption{Total simulation time and runtime memory scaling limits of different approaches for the Hubbard model with lattice size $N_s=4,8,12$. The KBE solver was accelerated with 48 OpenMP threads therefore it exhibits approximately an $O(N_t^{2.5})$ scaling with respect to the total number of timesteps $N_t$. The RNN solver exhibits a linear runtime and memory scaling, which is consistent with the theoretical analysis.}
\label{fig:runtime}
\end{figure}

\begin{table}
    \centering
       \begin{tabular}{ccccc}
$N_s$ & RNN & Storage & KBE & Storage  \\
\hline\hline
4 & 63.4s+16.9s &0.171 MB &127.041s & 0.686 GB \\
8 & 61.3s+54.2s & 0.684 MB & 416.883s & 2.744 GB\\
12 & 64.7s+116.7s &1.539 MB & 816.648s & 6.174 GB\\
24 & 89.2s+746.6s & 6.161 MB& 4516.47s & 24.694 GB\\
\end{tabular}
\caption{Scaling of runtime and memory consumption with respect to the Hubbard model system size $N_s$ for different approaches. The second column shows the required training $+$ dynamics extrapolation time for the RNN model (2 LSTM layers, 512 hidden states) to finish $5000$ training epochs with input/output data length $L=250$ and extrapolate up to $T=70$. The third column indicates the required runtime memory when performing dynamics extrapolation.  The fourth column shows the total simulation time to finish the same computational task using the NESSi package to solve KBEs but only up to $T=25$ ($N_t=1000$, 48 OpenMP threads) since longer-time data for $N_s=24$ are hard to obtain. The required memory for saving the Green's function and self-energy components in the two-time grid is displayed in the last column.}
\label{table:rnn_cost}
\end{table}
With the examples above, we have shown the capability of the operator-learning method in predicting the NEGF dynamics. As we mentioned in the introduction, the main advantage of this approach is that it reduces the computational cost for KBE from cubic scaling $O(N_t^3)$ to nearly linear scaling $O(N_t)$, which makes it possible to perform large-scale simulation for nonequilibrium many-body systems. Here we provide a detailed runtime analysis to clearly show the numerical scaling of the computational cost when applying this approach to different systems and highlight its advantages over existing methods. The numerical metrics for the test examples are summarised in TABLE \ref{table:rnn_cost} and the scaling limits are shown FIG \ref{fig:runtime}. 
The former illustrates the drastic computational saving brought by the dynamics reduction procedure since it enables performing one-time dynamics extrapolation in parallel that leads to the big reduction of the total simulation time and runtime memory. 
%
%
As for the operator-learning method we developed, the computational cost comes from two parts: 1. Training of the neural network and 2. Solving the EOM \eqref{EOM_G_subdiag} in parallel. From the second column of TABLE \ref{table:rnn_cost}, we see that the training time of the RNN grows extremely slowly with respect to the system size (slower than $O(N_s)$), which indicates a low training cost even for large systems. This slow growth is due to the RNN architecture (c.f. FIG \ref{fig:RNN_architecture}): In our construction, the MLP layer maps hidden states with fixed dimensions into the target time series. As $N_s$ increases, only the matrix dimensions in the MLP layer change with respect to $N_s$ hence the total number of modeling parameters in the neural network grows as of $O(N_s^2)$. For modern-day machine learning, it is considered a small optimization task for the two-band model where $N_s =24$, and one can readily generalized it to much larger systems with low computational cost. 

In the dynamics extrapolation phase, we see from FIG \ref{fig:runtime} that the computational cost of the prediction-correction scheme we used for solving EOM \eqref{EOM_G_subdiag} grows as of $O(N_t)$ while the KBE solver scales as of $O(N_t^{2.5})$ (the speedup from $O(N_t^3)$ to $O(N_t^{2.5})$ is achieved by parallelization). This is as expected because in essence \eqref{EOM_G_subdiag} is an ODE and the RNN generate outputs via function compositions therefore immediate. Also, since ODEs can be solved in a fine scale $dt$ but stored in a coarse-grained one-time grid $\Delta t$, the runtime memory consumption scales as of $O(N_s^2\tilde N_t)$ $\tilde N_t=\frac{dt}{\Delta t}N_t\ll N_t$, in contrast with the  $O(N_s^2 N_t^2)$ scaling of the KBE solver, as indicated in the subplot of FIG \ref{fig:runtime}.    
%
%
It is also noteworthy that the dynamics extrapolations for $G^<(t,t-a)$ with different $a$ are independent tasks that can be easily parallelized using MPI (c.f. Section \ref{sec:Parallelization}), the computational gain using the RNN approach for predicting the Green's function dynamics is obvious.  

%
%
\section{Conclusion}
From designing novel materials with tailored properties to developing advanced quantum computing architectures, the ability to perform comprehensive studies for nonequilibrium quantum many-body systems is profound and far-reaching. However, the intrinsic high-dimensionality of the problems in this field also presents formidable theoretical and computational challenges for modern material sciences and applied mathematics. In this work, we proposed an RNN-based nonlinear operator learning framework to learn and predict the dynamics of NEGFs. By integrating the machine-learned collision integral operator with the mean-field solver of KBEs and then using the dynamics reduction techniques, eventually, we used RNN with relatively simple architectures and low training costs to realize the dynamics learning/predicting for a system of high-dimensional nonlinear integro-differential equations of four-rank tensors i.e. NEGFs. The final computational cost was successfully reduced from $O(N_t^3)$ scaling to $O(N_t)$ with runtime memory consumption dropped down from $O(N_t^2N_s^2)$ to $O(\tilde N_tN_s^2)$, where $\tilde N_t\ll N_t$. The proposed algorithm has been tested in many different nonequilibrium quantum many-body systems, and it showed excellent accuracy, numerical convergence, remarkable scalability, and is also easy to parallelize. Comparing to other linearly scaling solvers such as the HF-GKBA \cite{joost2020g1}, our approach does not increase the spatial complexity of the calculation since it avoids the two-electron correlator calculations. Moreover, in the time subdiagonal region, the inclusion of the collision integral $I^<(t,t+a)$ yields more accurate prediction of the Green's function dynamics, hence of the photoemission spectra. 

Our work laid the groundwork for many potential extensions and subsequent investigations. From a methodological standpoint, the versatility of machine learning architectures allows for various adaptations, including the incorporation of attention-based frameworks \cite{vaswani2017attention} for dynamics learning and extrapolation. More importantly, as an operator-learning approach, the new framework has natural generalizability and one could use the learned collision integral operator to make dynamics predictions for systems driven by different nonequilibrium forces. In the end, the remarkable scalability of our approach made it possible to perform long-time simulations for realistic nonequilibrium systems where direct simulation results could not be achieved. This potential suggests promising avenues for further exploration and application in condensed matter physics and relevant fields.  
\section{Numerical simulation details}\label{sec:Methods}
%
%
%
%

\subsection{Data source}
The NEGFs are obtained by solving the KBEs using the NESSi library \cite{Schuler_2020} with the second-order Born self-energy (c.f. Supplementary Note Section 1). The time integrator stepsize is set to be $0.025J^{-1}$. A large enough inverse temperature $\beta=20$ is chosen for simulating zero-temperature dynamics. In this work, we use NN to learn the mapping between the lesser Green's function $G^<(t,t')$ and the lesser collision integral $I^<(t,t')$. This implicitly assumes that the dynamics of $G^<(t,t')$ is self-consistent, although the KBEs indicate that it should also depend on other Keldysh components such as the greater Green's function $G^R(t,t')$ (c.f. Supplementary Note Section 1). As a ML method, this simplification reduces the dimensionality of the optimization problem. Mathematically, it is also reasonable since one can always formally express $G^R(t,t')$ as functional of $G^<(t,t')$. All the complexity is hidden in the formal mapping $G^<(t,t')\rightarrow I^<(t,t')$, which is assumed can be learned by RNN. 

\subsection{Learning and Training details}
\label{subsec:Learning and Training details}
The RNN architecture we used was indicated in FIG \ref{fig:RNN_architecture}. For all training tasks, we use an RNN with 2 LSTM layers with hidden size 512 to learn/extrapolate the Green's function dynamics. The parameters are optimized with Adams' optimizer with learning rate $10^{-3}$ after 5000 training epochs. The input/output of the neural network is the {\em time-coarse-grained} Green's function data. Specifically, the input time sequence is $[G(0),G(\Delta t),\cdots,G(T)]$ with $\Delta t= 0.1$, and so is the output time sequence. Each $G(i\Delta t)$ is a $2N_s^2$-dimensional real-valued vector, which is obtained by splitting the real and imaginary part for each matrix element $G_{ij}(t)$ and flattening the matrix into vectors. The time-coarse-graining of the data is mainly for improving the convergence rate and the accuracy of the numerical optimization. $G(i\Delta t)$ is flattened into a real-valued vector so that we can conveniently implement the RNN using machine learning libraries such as Pytorch.    

In contrast with our prior study \cite{bassi2024learning}, an important adaption here is the reduction of the dynamics of two-time Green's function $G^<(t,t')$ into one-time, following the procedure outline in FIG \ref{fig:dynamics_reduction}. Specifically, we first learn the time-diagonal dynamics $G^<(t,t)$, where the input/output of the RNN is $G^<(t,t)$ and $I^<(t,t)$ (Region I in FIG \ref{fig:dynamics_reduction}). Then we learn the time-subdiagonal dynamics $G^<(t,t-a)$ for different $a$, where the input/output of the RNN becomes $G^<(t,t-a)$ and $I^<(t,t-a)$ (Region II in FIG \ref{fig:dynamics_reduction}). Lastly, the time-offdiagonal dynamics $G^<(t,b)$ for different $b$ can be calculated in a similar way (Region III in FIG \ref{fig:dynamics_reduction}). In the paper, we only show numerical results for the first two steps because the RNN-learned nonlinear mapping $G^<(t,b)\rightarrow I^<(t,b)$ is found to have limited predictability of future time dynamics for $G^<(t,b)$ (See numerical results in Supplementary Note Section 2.3), which is in contrast with the diagonal and subdiagonal cases. We believe this is related to the strong memory effect of the off-diagonal dynamics. Fortunately, for calculating important physical observables such as the one-particle reduced density matrix and photoemission spectra, it is often sufficient to know only the time-diagonal and subdiagonal dynamics as we have discussed in Section \ref{sec:Results}. 

\subsection{Extrapolation}
\label{sec:extrapolation}
After the RNN is trained after a certain amount of optimization epochs, or when the target error bound is achieved, the neural network can be used to predict the Green's function dynamics following the procedure outlined in FIG \ref{fig:RNN_extrapolation}. Since the RNN is trained using the one-time data, the dynamics extrapolation has to be done along the time-diagonal and time-subdiagonals of the Green's function. From the KBE \eqref{KBE}, we could extract the reduced EOM for $G^<(t,t)$ and $G^<(t,t-a)$ (c.f. Supplementary Note Section 1):
\begin{align}
i\partial_t G^<(t,t)&=[h^{\textrm{HF}}(t),G^<(t,t)] +\hat I^<(t,t)\label{EOM_G_diag}\\
i\partial_t G^<(t,t-a)&=h^{\textrm{HF}}(t)G^<(t,t-a)\nonumber\\
&-G^<(t,t-a)h^{\textrm{HF}}(t-a) + \hat I^<(t,t-a).
\label{EOM_G_subdiag}
\end{align}
Here we note that $h^{\textrm{HF}}(t)=h^{\textrm{HF}}(G^<(t,t),t)$, therefore one needs to know the time-diagonal data $G^<(t,t)$ before soling for time-subdiagonal Green's function $G^<(t,t-a)$. As a result, we have to first solve the self-consistent EOM \eqref{EOM_G_diag} to get predicated $G^<(t,t)$ and then using the obtained $G^<(t,t)$ to solve \eqref{EOM_G_subdiag} for $G^<(t,t-a)$ with different $a$. For TDHF solver, we simply set $\hat I^<$ to be zero in both equations and use the 5th-order Adams–Bashforth (AB5) scheme with stepsize $dt=0.005$ to perform numerical integration. For RNN-based solver, the numerical integrator we used is a prediction-correction scheme based on AB5. The prediction-correction procedure is needed since the input/output of the RNN is time-coarse-grained data with stepsize $\Delta t=0.1$, which is too large for accurately simulating the Green's function dynamics. Hence, in each timestep, we first solve \eqref{EOM_G_diag} and \eqref{EOM_G_subdiag} using AB5 with stepsize $\Delta t=0.1$ to get $G(T+\Delta t)$. Then it is fed into the RNN to generate a {\em predicted} collision integral $\hat I(T+\Delta t)$. Then we use cubic-spline interpolation to get the fine-scale data $[\hat I(T),\hat I(T+dt),\cdots,\hat I(T+\Delta t)]$, where $dt=0.005$. In the end, the {\em corrected} Green's function $G(T+\Delta t)$ and $\hat I(T+\Delta t)$ is obtained by solving the same EOM using AB5 with stepsize $dt=0.005$ in the fine time-grid $[T,T+dt,\cdots, T+\Delta t]$. 

When we learn the integral operator mapping in time-subdiagonals, i.e. $G^<(t,t-a)\rightarrow I^<(t,t-a)$, the RNN will get lesser training data for as $a\rightarrow T$, where $T$ is the training data length (c.f. FIG \ref{fig:dynamics_reduction}). This data inadequacy naturally leads to ineffective operator-learning and therefore wrong predictions of the future time-dynamics. When $a$ is bigger than $T/2$, we found that the simple mean-field (HF) extrapolation will normally yield a stable and accurate prediction of the future time dynamics of $G^<(t,t-a)$. Of course, one can also use the adaptive procedure we outline in Supplementary Note Section 3 to automatically determine the range of $a$ that yield accurate dynamics predictions. In practice, for calculating the photoemission spectra, this limitation is not severe since the time-subdiagonal data are are close to the time-diagonals contributes more to the final calculation result of the spectral function, as we have seen in Section \ref{sec:emission_spectra}. 

\subsection{Parallelization}
\label{sec:Parallelization}
The dynamics reduction procedure enables a simple parallelization for learning and extrapolating the time-subdiagonal Green's function $G^<(t,t-a)$. Namely, for different $a$, the EOM \eqref{EOM_G_subdiag} are independent of each other. Therefore one can easily distribute the job to different MPI ranks and do computations in parallel as there is no communication between them.

\section{Acknowledgement}
This material is based upon work supported by the U.S. Department of Energy, Office of Science, Office of Advanced Scientific Computing Research and Office of Basic Energy Sciences, Scientific Discovery through Advanced Computing (SciDAC) program under Award Number DE-SC0022198.  This research used resources of the National Energy Research Scientific Computing Center, a DOE Office of Science User Facility supported by the Office of Science of the U.S. Department of Energy under Contract No. DE-AC02-05CH11231 using NERSC award BES-ERCAP0029462 (project m4022) and ASCR-ERCAP-m1027.

\section{Data and code availability}
If accepted, all data and code will be shared via an open source link.

\section{Ethical Statement}
The research presented in this article does not involve human participants, animals, or any data or materials that require ethical approval. Therefore, no ethical issues are associated with this work. All research was conducted in accordance with the relevant national and international standards for academic integrity and ethical conduct.

\section{Competing interests}
The authors declare no competing interests.

%
%
    
\bibliography{apssamp}
\end{document}


\title{Supplementary Note}
\maketitle

\section{Kadanoff-Baym equation and self-energy approximation}\label{app:sec1}
The Kadanoff-Baym equation is a system of integro-differential equations defined in the Keldysh contour $\mathcal{C}$ of the complex plane, where $\mathcal{C}=\{z\in\mathbb{C}|\mathrm{Re}[z]\in[0,+\infty],\mathrm{Im}[z]\in[0,-\beta]\}$ \cite{stefanucci2013nonequilibrium}. The EOM is given by:
\begin{equation}\label{KBE_formal}
\begin{aligned}
\left[i\partial_z - h(z)\right]G(z,z') &= \delta(z,z')  + \int_{\mathcal{C}}\mathrm{d}\bar{z} \Sigma(z,\bar{z})G(\bar{z},z')\\
\left[-i\partial_{z'} - h(z)\right]G(z,z') &= \delta(z,z')  + \int_{\mathcal{C}}\mathrm{d}\bar{z} G(z,\bar{z})\Sigma(\bar{z},z')
\end{aligned}
\end{equation}
for $z,z'\in\mathcal{C}$. Introducing a set of  Green's functions $G^{\lessgtr}(z,z'),G^{\rceil/\lceil}(z,z'), G^M(z)$ which are defined in different branches of the contour, the KBE can be reformulated into an equivalent system of integro-differential equations given by \cite{balzer2012nonequilibrium}: 
\begin{equation}\label{eq:KBE}
           \begin{split}
                [-\partial_\tau - h] G^\mathrm{M}(\tau) &= \delta(\tau) + \int_0^\beta d\Bar{\tau}\Sigma^\mathrm{M}(\tau-\Bar{\tau})G^\mathrm{M}(\bar{\tau}),\\
                i\partial_{t_1} G^{\rceil}(t_1,-i\tau) &= h^{\textrm{HF}}(t_1)G^{\rceil}(t_1,-i\tau) + I^{\rceil}(t_1,-i\tau),\\
                -i\partial_{t_2} G^{\lceil}(-i\tau,t_2) &= G^{\lceil}(-i\tau,t_2)h^{\textrm{HF}}(t_2) + I^{\lceil}(-i\tau,t_1),\\
                i\partial_{t_1} G^{\lessgtr}(t_1,t_2) &= h^{\textrm{HF}}(t)G^{\lessgtr}(t_1,t_2) + I_1^{\lessgtr}(t_1,t_2),\\
                -i\partial_{t_2} G^{\lessgtr}(t_1,t_2) &= G^{\lessgtr}(t_1,t_2)h^{\textrm{HF}}(t_2) + I_2^{\lessgtr}(t_1,t_2),\\
            \end{split}
\end{equation}
where $0\leq t_1,t_2<+\infty$, $\tau\in[0,\beta]$, and $I(t,t')$ are collision integrals in different branches explicitly given by \cite{balzer2012nonequilibrium}:
\begin{equation}\label{eq:coll_int}
    \begin{split}
        I_{1}^{\lessgtr}(t_1,t_2) &= \int_{0}^{t_1}\mathrm{d}\bar{t} \Sigma^\mathrm{R}(t_1,\Bar{t})G^{\lessgtr}(\Bar{t},t_2) +\int_{0}^{t_2} \mathrm{d}\bar{t} \Sigma^{\lessgtr}(t_1,\Bar{t})G^\mathrm{A}(\Bar{t},t_2)- i\int_0^\beta \mathrm{d}\bar{\tau} \Sigma^\rceil(t_1,-i\bar{\tau})G^{\lceil}(-i\bar{\tau},t_2),\\
        I_{2}^{\lessgtr}(t_1,t_2) &= \int_{0}^{t_1} \mathrm{d}\bar{t} G^\mathrm{R}(t_1,\Bar{t})\Sigma^{\lessgtr}(\Bar{t},t_2) + \int_{0}^{t_2} \mathrm{d}\bar{t} G^{\lessgtr}(t_1,\Bar{t})\Sigma^\mathrm{A}(\Bar{t},t_2) - i\int_0^{\beta}\mathrm{d}\bar{\tau}G^{\rceil}(t_1,-i\bar{\tau})\Sigma^{\lceil}(-i\bar{\tau},t_2),\\
        I^{\rceil}(t_1,-i\tau) &= \int_{0}^{t_1} d\bar{t} \Sigma^\mathrm{R}(t_1,\bar{t})G^{\rceil}(\bar{t},-i\tau) + \int_0^\beta d\bar{\tau} \Sigma^{\rceil}(t_1,-i\bar{\tau})G^M(\bar{\tau} - \tau),\\
        I^{\lceil}(-i\tau,t_1) &= \int_0^{t_1}d\bar{t} G^{\lceil}(-i\tau,\bar{t})\Sigma^\mathrm{A}(\bar{t},t) +\int_0^{\beta}d\bar{\tau}G^\mathrm{M}(\tau - \bar{\tau}) \Sigma^{\lceil}(-i\bar{\tau},t_1).
    \end{split}
\end{equation}
The self-energy terms $\Sigma^{R/A}(t,t'),\Sigma^{\rceil}(t,-i
\tau),\Sigma^{\lceil}(-i
\tau,t),\Sigma^M(\tau)$ (beyond Hartree-Fock) take into account the many-body correlation of the system. In this paper, we use the second-order Born (2ndB) approximation for $\Sigma$. For the paramagnetic Hubbard model at half-filling, we have:
\begin{align*}
h^{HF}_{ij}(t)&=h_{ij}(t)-iUG_{ii}^{<}(t,t)\\
\Sigma^{2ndB}_{ij}(t,t')&= U^2G_{ij}(t,t')G_{ij}(t,t')G_{ji}(t',t),
\end{align*}
where the explicit expression for $\Sigma^{R/A}(t,t'),\Sigma^{\rceil}(t,-i
\tau),\Sigma^{\lceil}(-i
\tau,t),\Sigma^M(\tau)$ follows from the Langreth rule and can be found in \cite{balzer2012nonequilibrium,Schuler_2020,stefanucci2013nonequilibrium}. From the lesser Green's function $G^<(t_1,t_2)$, choosing $t_1=t_2=t$, $t_1=t,t_1=t-a$, and $t_1=t,t_2=b$ respectively, one can obtain the following reduced EOMs for different one-time Green's functions:
\begin{equation}
\begin{aligned}
i\partial_t G^<(t,t)&=[h^{HF}(t),G^<(t,t)] +I^<(t,t)\\
i\partial_t G^<(t,t-a)&=h^{HF}(t)G^<(t,t-a)-G^<(t,t-a)h^{HF}(t-a) + I^<(t,t-a)\\
i\partial_t G^<(t,b)&=h^{HF}(t)G^<(t,b)+ I^<(t,b).
\end{aligned}
\end{equation}
\begin{figure}[h]
\centering
\includegraphics[width=17cm]{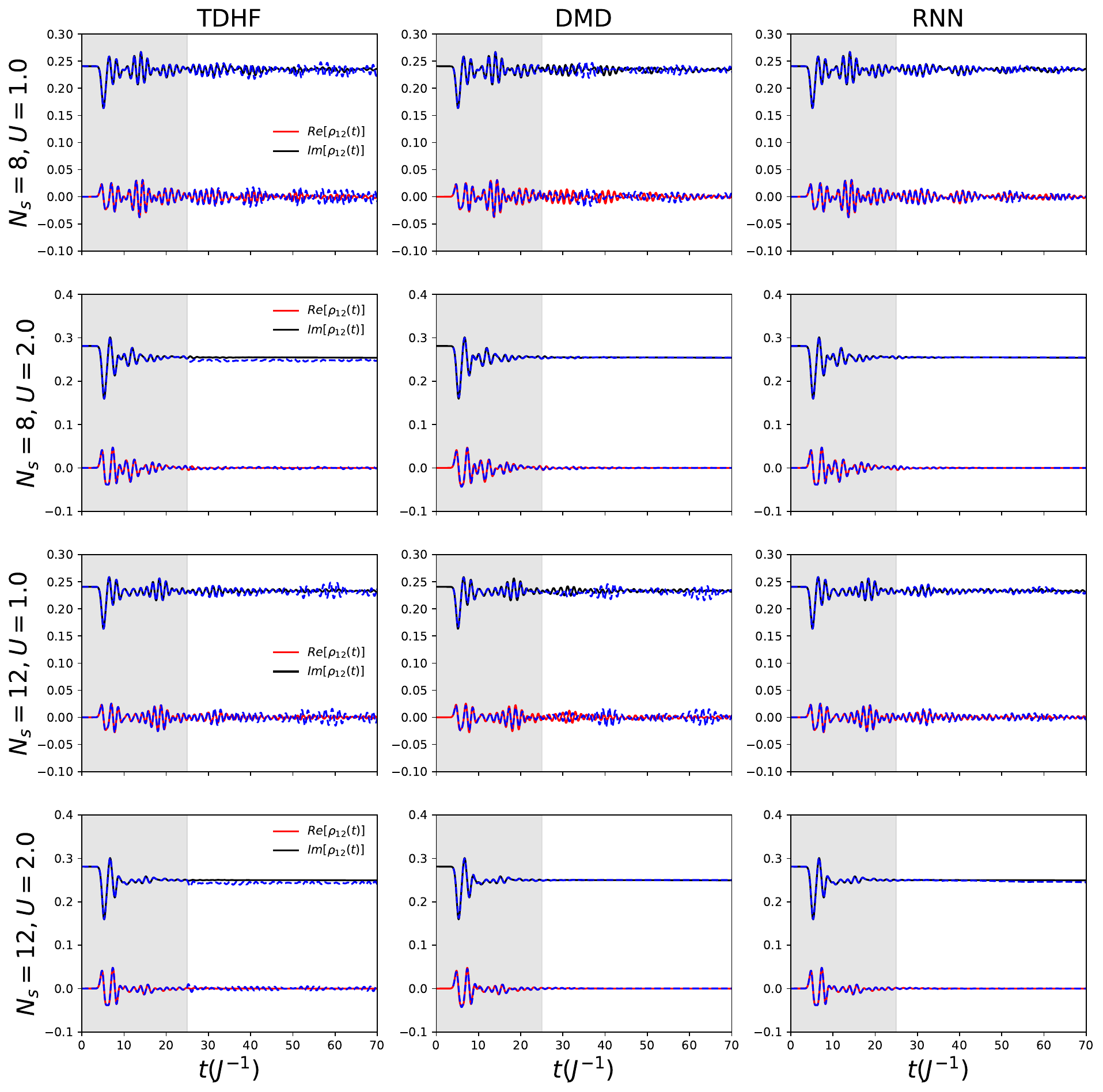}
\captionsetup{justification=justified}
\caption{Comparison of the predicted 1RDM $\rho(t)=-iG^<(t,t)$ generated by different methods. For the DMD and RNN approach, the KBE solution up to $T=25$ (indicated by the shaded region) was used as the learning data. In all figures, the dashed line displays the approximated dynamics and the solid line indicates the KBE result. The tested model is the half-filled Hubbard model driven by the long wavelength force with modeling parameters $\{J=1,U=1.0,2.0,V=2,N_s=8,12,E=1.0, T_p= 0.5\}$ at the inverse temperature $\beta =20$.}
\label{fig:Gles_diag_longwave}
\end{figure}
\section{Dynamics prediction by different approaches}
In this section, we provide more numerical examples to show the dynamics predictability of the RNN for different nonequilibrium many-body systems. We also compare the predicted Green's function generated by our approach with what obtained by the TDHF solver and DMD method. Here TDHF provides a convolution-less baseline that highlights the memory effect of the system. It can show how well the RNN learned the collision integral operator. DMD method has been proven to be successful in predicting NEGF dynamics \cite{yin2022using,yin2023analyzing}. As a representative data-driven method, it is constructive to make comparison between results generated by DMD and RNN method.  
\begin{figure}[h]
\centering
\includegraphics[width=17cm]{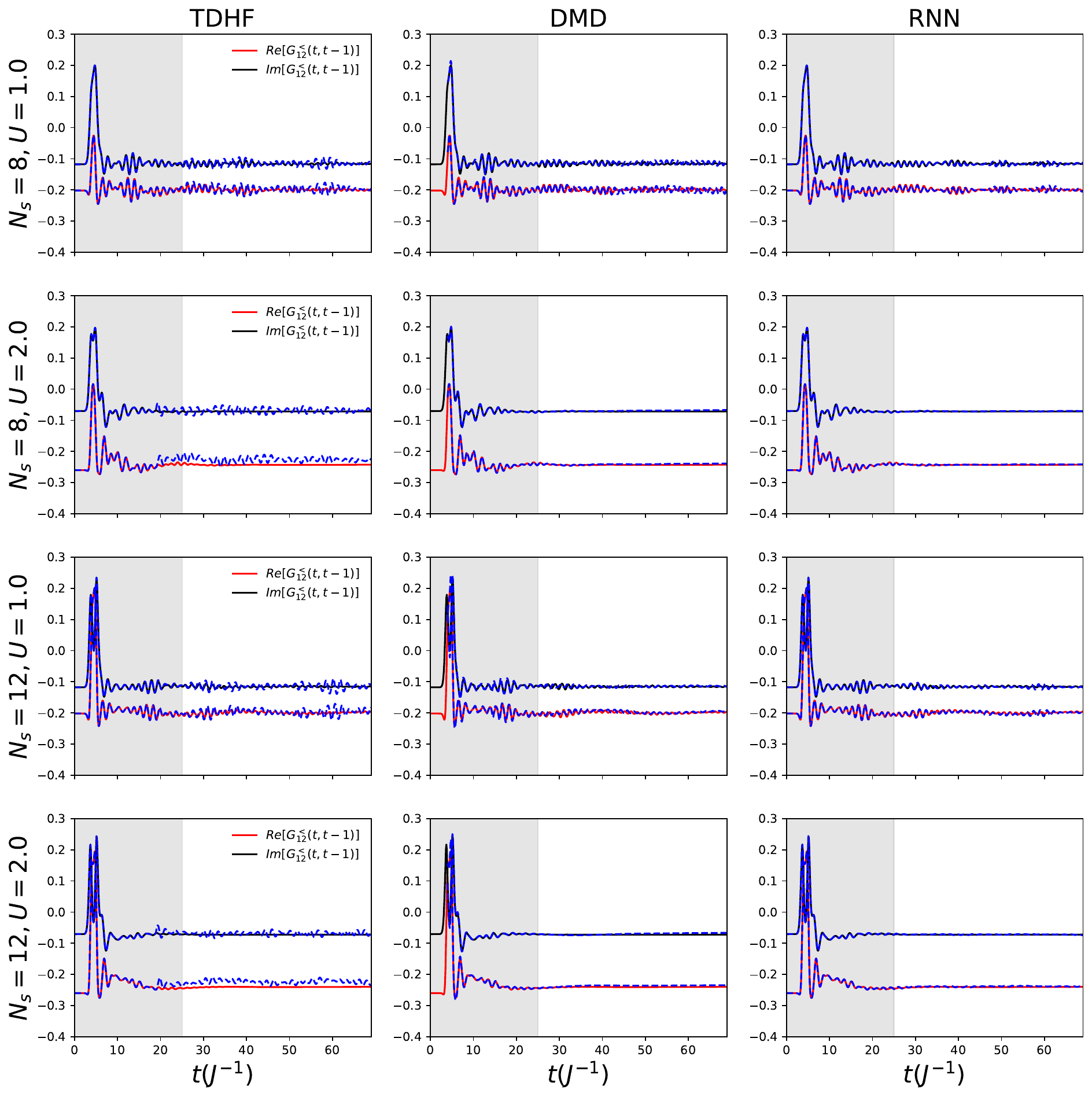}
\captionsetup{justification=justified}
\caption{Comparison of the time-subdiagonal Green's function $G^<(t,t-1)$ obtained by different dynamics extrapolation methods. The modeling parameters are the same as in Figure \ref{fig:Gles_diag_longwave}.}
\label{fig:Gles_subdiag_longwave}
\end{figure}
\begin{figure}[h]
\centering
\includegraphics[width=\textwidth]{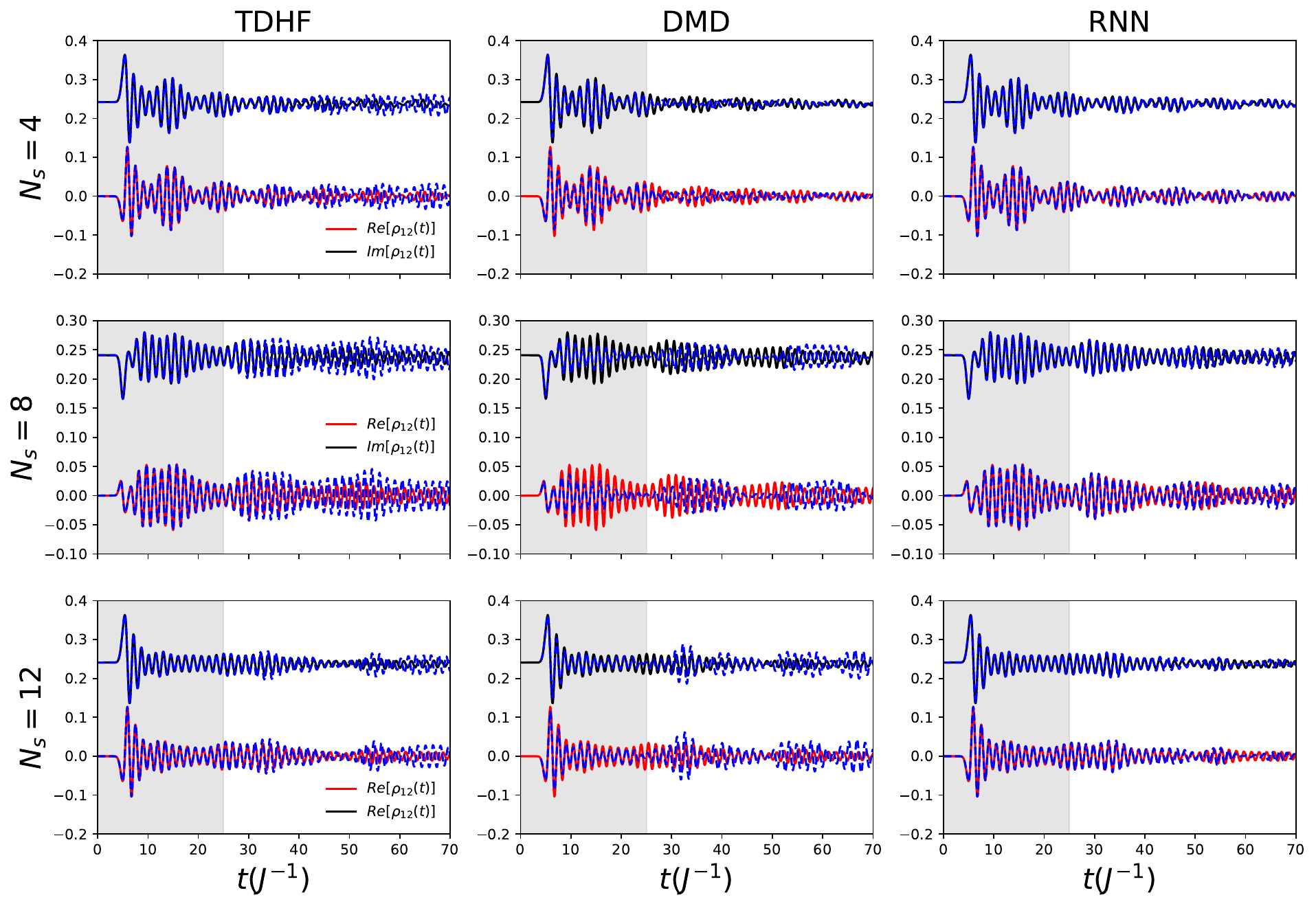}
\caption{Comparison of the predicated 1RDM $\rho(t)=-iG^<(t,t)$ generated by different methods. For the DMD and RNN approach, the KBE data up to $T=25$ (indicated by the shaded region) was used as the learning data. The tested model is the half-filled Hubbard model driven by the short wavelength force with modeling parameters $\{J=1,U=1.0,V=2,N_s=4,8,12,E=1.0, T_p= 0.5\}$ at the inverse temperature $\beta =20$.}
\label{fig:Gles_diag_shortwave}
\end{figure}
\subsection{Systems with different sizes and correlation strength}
We first compare the predicted lesser Green's function for the Hubbard model with different sites $N_s=8,12$ and interaction strength $U=1.0,2.0$. The results are shown in Figure \ref{fig:Gles_diag_longwave} and \ref{fig:Gles_subdiag_longwave}. As we can see, based on the KBE training data up to $T=25$, for all the test cases, the RNN approach provides reliable predictions of the Green's function dynamics in both the time-diagonal $\rho(t)=-iG^<(t,t)$ and time-subdiagonal $G^<(t,t-a)$ regions. By comparing with the TDHF result, we found the RNN result is more accurate. This is not surprising since the RNN-based KBE solver for Green's function incorporates the approximated collision integral, which is completely ignored in the TDHF solver. As the correlation strength gets larger to $U=2.0$ where the many-body correlation gets more prominent, the difference between the RNN and TDHF result becomes more obvious. 

For comparison with the DMD result, we choose the high-order DMD (HODMD) method \cite{yin2022using, reeves2023dynamic,mejia2023stochastic} to extrapolate the Green's function dynamics. 
This approach has been proved to be quite effective in many other applications \cite{yin2022using,yin2023analyzing,maliyov2023dynamic}. In specific implementation, we choose the order of HODMD and cut-off singular values case by case to make sure there is a small fitting error between the DMD approximates and the KBE solution within the training window. For some test examples like the $U=2.0$ cases, we see that the DMD method also provides a reasonable prediction of the Green's function. However, it appears to be less accurate and less consistent than RNN. We believe the discrepancy here is due to the lack of training data. In other words, for DMD to make a reasonable prediction of Green's function dynamics, in general, it requires more training data than RNN.  

\subsection{Short-wave perturbation force}
\begin{figure}[t]
\centering
\includegraphics[width=\textwidth]{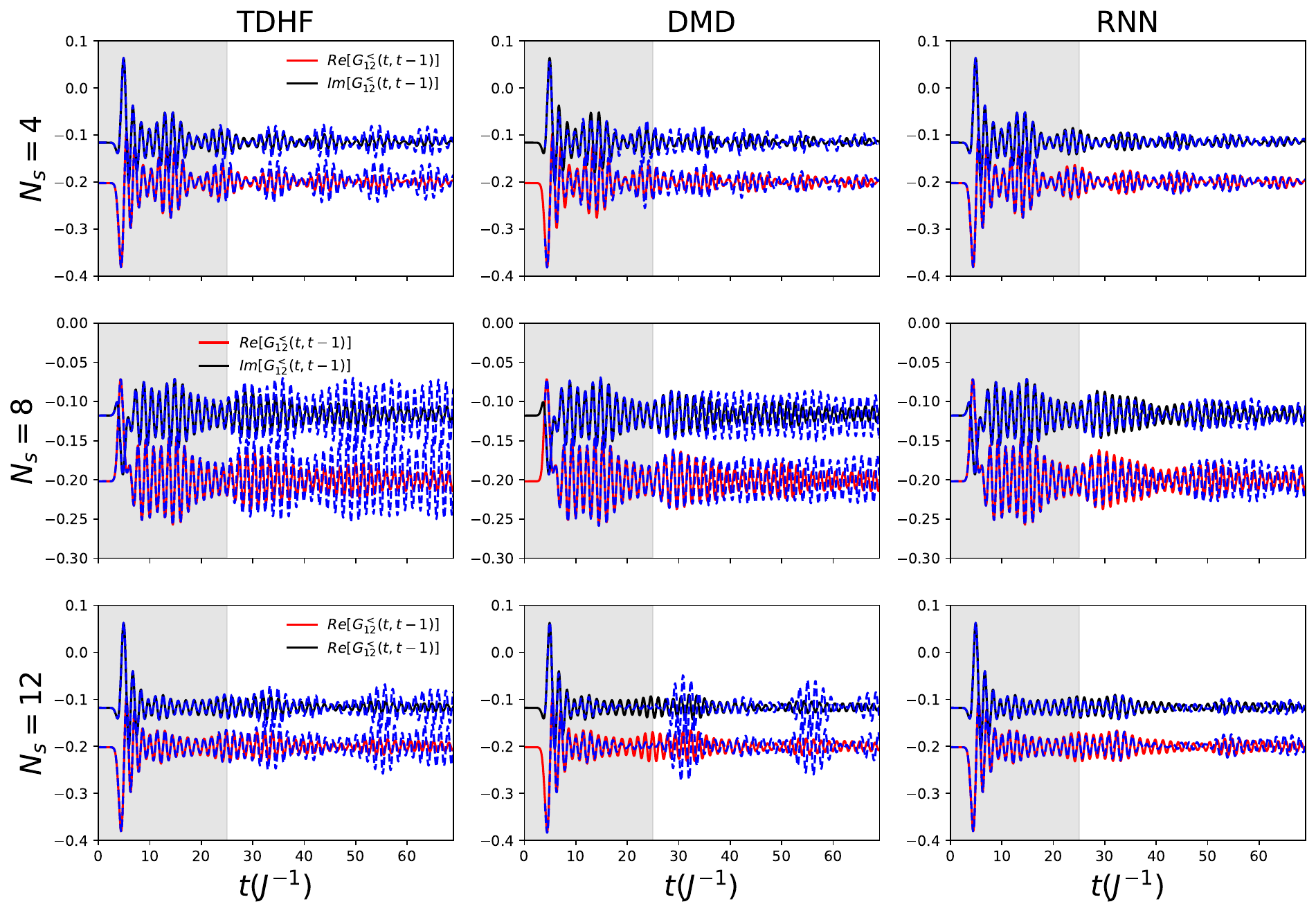}
\captionsetup{justification=justified}
\caption{Comparison of the extrapolated time-subdiagonal Green's function $G^<(t,t-1)$ by different dynamics extrapolation methods. Other settings are the same as in Figure \ref{fig:Gles_diag_shortwave}.}
\label{fig:Gles_subdiag_shortwave}
\end{figure}

The numerical examples in the last section are for Hubbard models driven by the long wavelength force. Here we provide similar test results for Hubbard models driven by the short wavelength force, which seems to provide more oscillations in the Green's function dynamics. The time-diagonal and time-subdigonal dynamics extrapolation results are summarized in Figure \ref{fig:Gles_diag_shortwave} and \ref{fig:Gles_subdiag_shortwave} respectively. As we can see, the RNN approach provides good predictions of the future-time dynamics for all test cases. In general, it is also more accurate than the TDHF and DMD approach. 

%
%
%
%
\subsection{Off-diagonal extrapolation}
\begin{figure}[t]
\centering
\includegraphics[width=\textwidth]{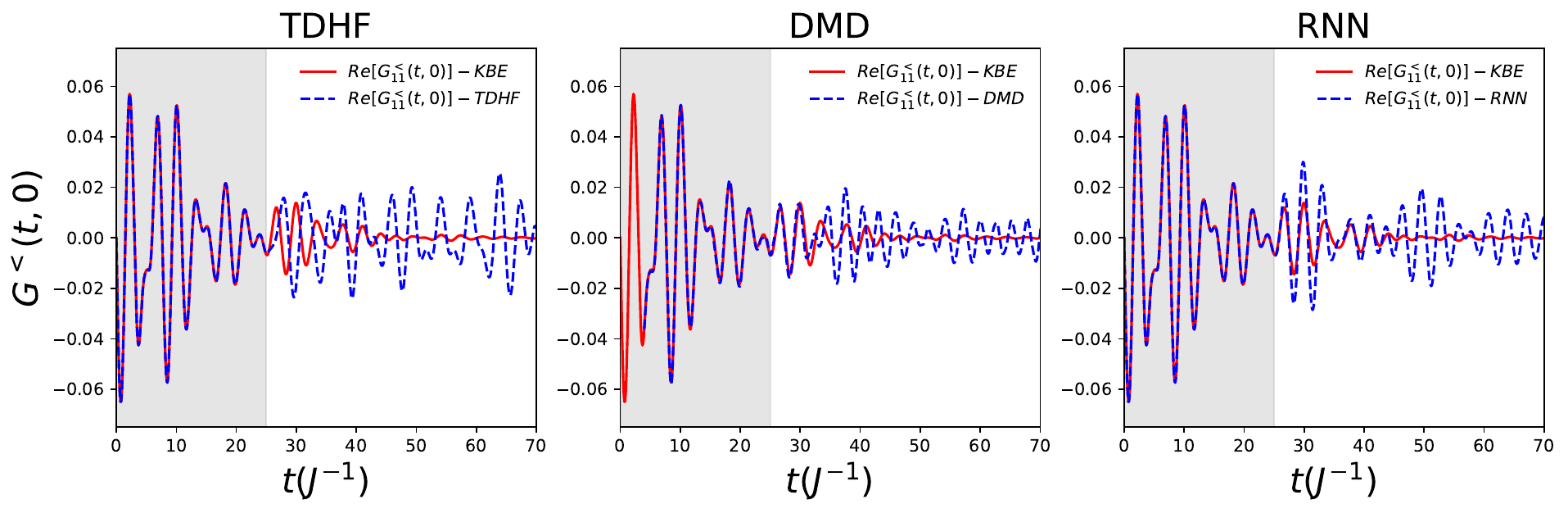}
\captionsetup{justification=justified}
\caption{Comparison of the extrapolated time-offdiagonal Green's function $G^<(t,0)$ by different dynamics extrapolation methods. The tested model is the 8-site half-filled Hubbard model driven by the long wavelength force with modeling parameters $\{J=1,U=1.0,V=2,N_s=8,E=1.0, T_p= 0.5\}$ at the inverse temperature $\beta =20$.}
\label{fig:Gles_offdiag_shortwave}
\end{figure}

For most applications, knowing the time-diagonal and time-subdiagonal Green's function would be enough to extract important physical information such as the occupation number and photoemission spectra from the underlying nonequilibrium dynamics. For the completion of our research, it is worthy to also perform numerical tests for the time-offdiaginal dynamics, i.e. for learning/predicting the dynamics of $G^<(t,a)$ where $a$ is fixed. In Figure \ref{fig:Gles_offdiag_shortwave}, we provide the dynamics extrapolation result of $G^<(t,0)$ for the Hubbard model with the same amount of training data (up to $T=25$). From the figure, we see that even for relatively small interaction strength $U=1.0$ where the electron correlation is not strong, the TDHF and RNN provide less accurate dynamics predictions in contrast with the time-diagonal and time-subdiagnal cases. The DMD prediction is not accurate either. We believe this attributes to the fact that the system's correlation effect is more prominent along the time-offdigonals. A very obvious numerical observation that supports this claim is that for the case of $U=1.0$, the matrix elements of the collision integrals $I^<(t,t),I^<(t,t-a)$ are in general two-magnitude smaller than that of Green's functions (similar observations are also reported in our previous work \cite{reeves2023unimportance} for many different systems). While for $I^<(t,a)$, they are the same order as of $G^<(t,a)$. 
%
%
\section{Numerical convergence analysis}
For data-driven methods to make dynamics prediction, the future dynamics of the underlying system are unknown, hence an obvious technical difficulty is to {\em predetermine} how much training data is enough to make a reasonable prediction of the dynamics. This problem is hard to address from a theoretically point of view, especially for large-scale complex dynamical systems like KBEs. However, we could design numerical criteria for the convergence assessments which enable us to provide an adaptive procedure to predetermine the required training data for dynamics extrapolations.         

\begin{figure}
\centering
\begin{subfigure}{\textwidth}
\centering
\includegraphics[width=14cm]{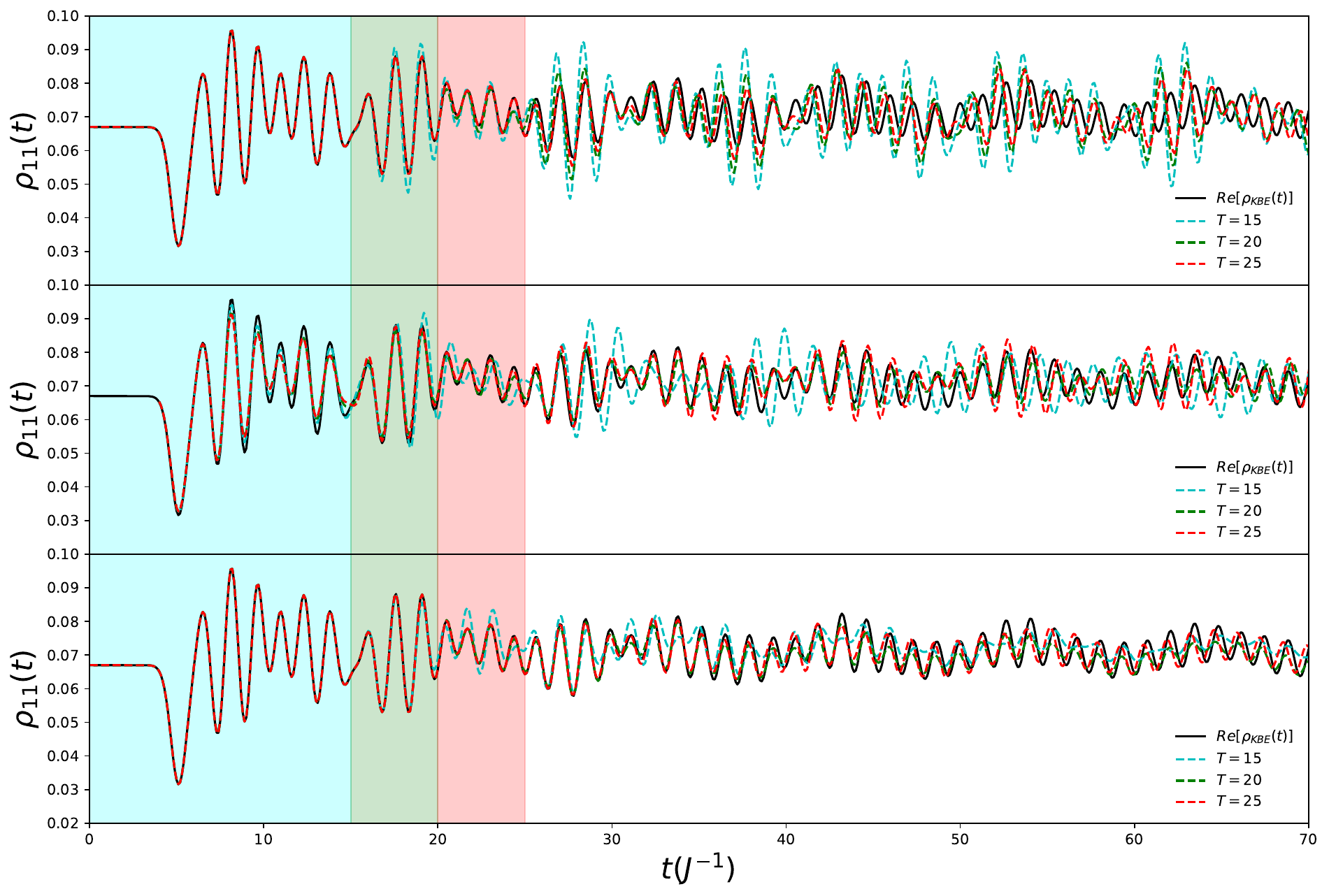}
\end{subfigure}
\caption{Numerical convergence test for three dynamics extrapolation methods: TDHF (first row), DMD (second row) and the RNN (third row). Each subfigure compares the extrapolated $\rho(t)$ with the KBE solution as we gradually increase the training data length $T=15,20,25$. The test model is the 4-site half-filled Hubbard model driven by the long wavelength force with modeling parameters $\{J=1,U=1/2,V=2,N_s=4,E=1.0, T_p= 0.5\}$ at the inverse temperature $\beta =20$.}
\label{fig:convergence_test_N4}
\vspace{0.5cm}
\centering
\begin{subfigure}{\textwidth}
\centering
\includegraphics[width=14cm]{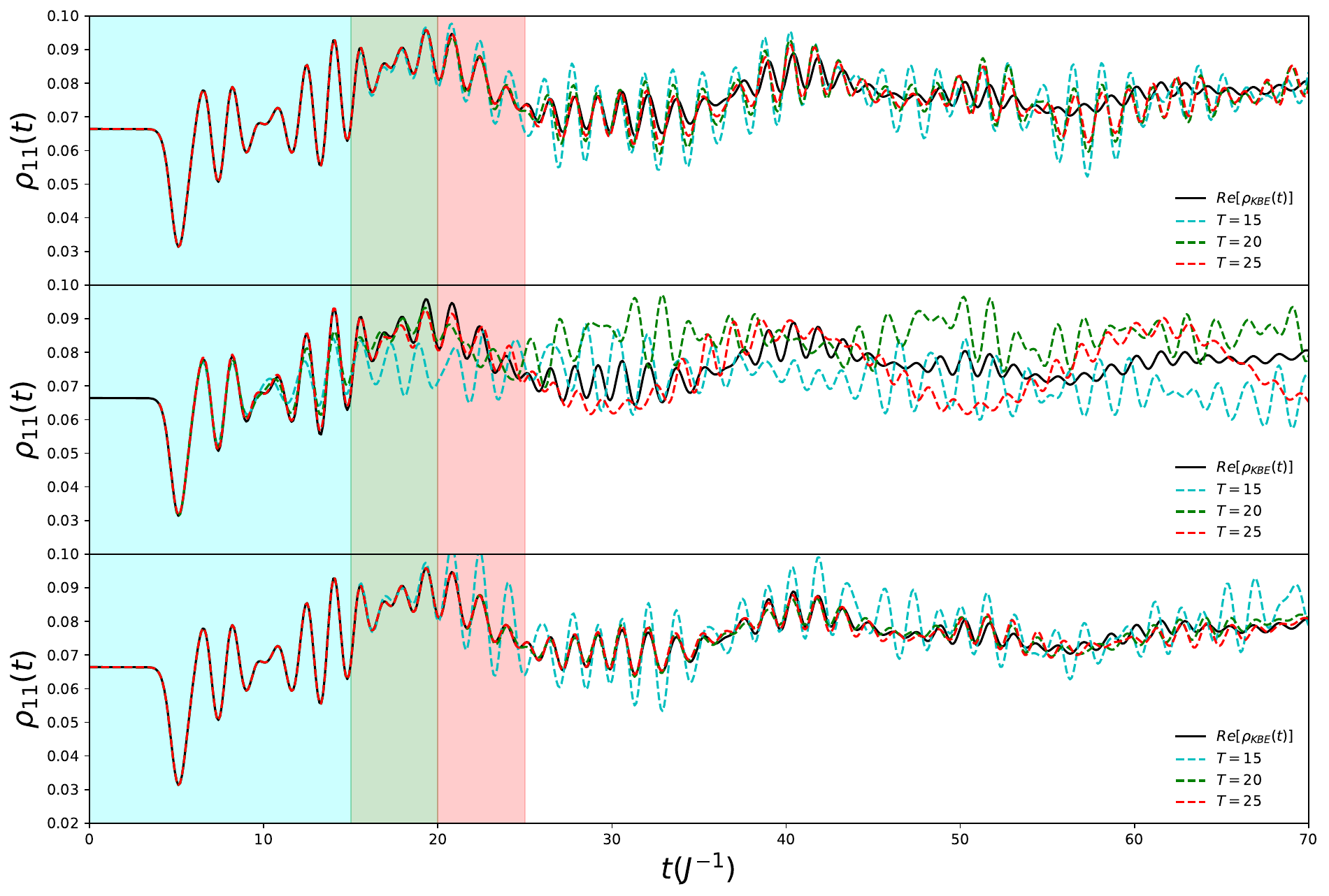}
\end{subfigure}
\captionsetup{justification=justified}
\caption{Numerical convergence test the 8-site half-filled Hubbard model driven by the long wavelength force with modeling parameters $\{J=1,U=1.0,V=2,N_s=8,E=1.0, T_p= 0.5\}$ at the inverse temperature $\beta =20$.}
\label{fig:convergence_test_N8}
\end{figure}

\begin{figure}[h]
\centering
\begin{subfigure}{\textwidth}
\centering
\includegraphics[width=15cm]{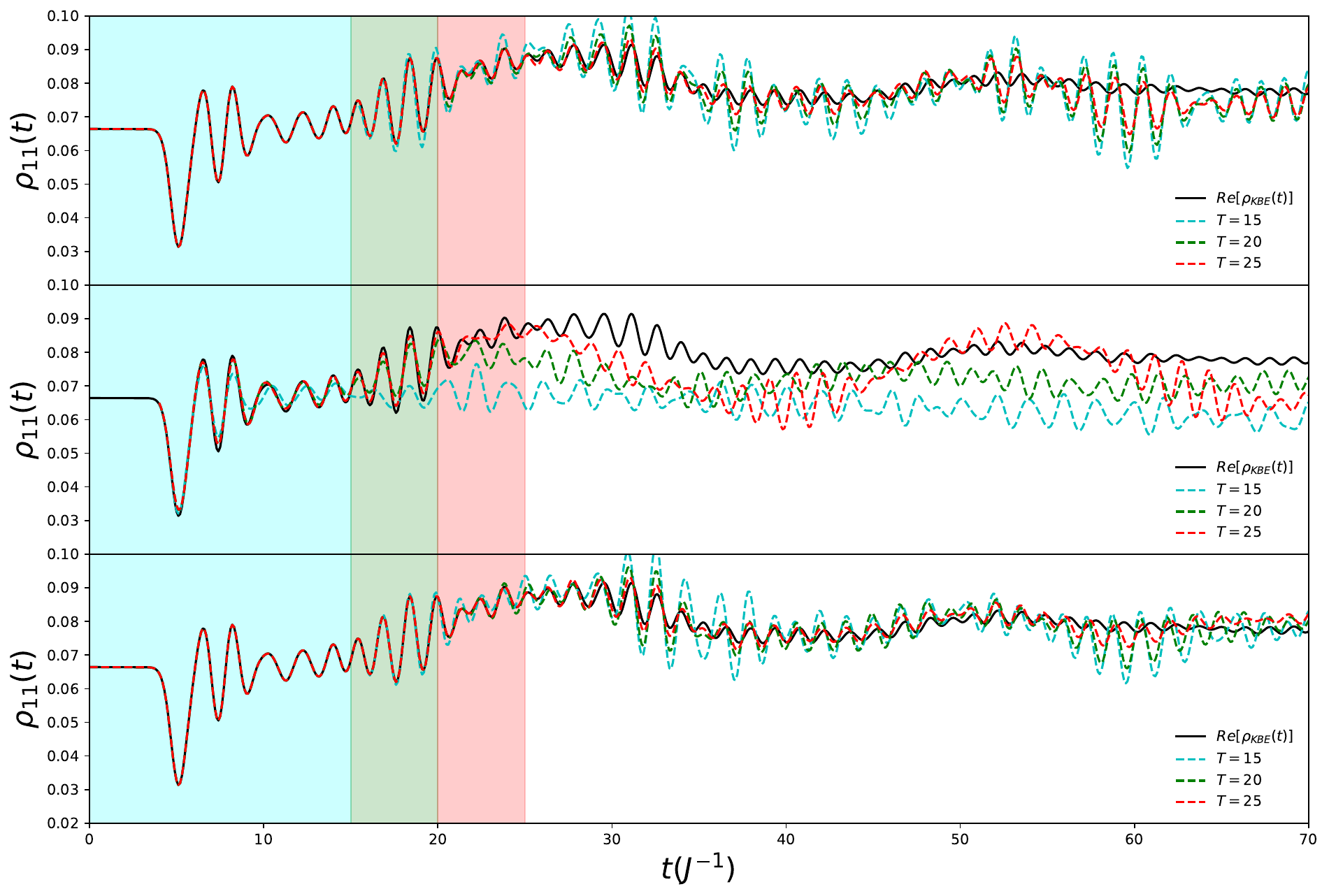}
\end{subfigure}
\caption{Numerical convergence test the 12-site half-filled Hubbard model driven by the long wavelength force with modeling parameters $\{J=1,U=1.0,V=2,N_s=12,E=1.0, T_p= 0.5\}$ at the inverse temperature $\beta =20$.}
\label{fig:convergence_test_N12}
\centering
\begin{subfigure}{\textwidth}
\centering
\includegraphics[width=15cm]{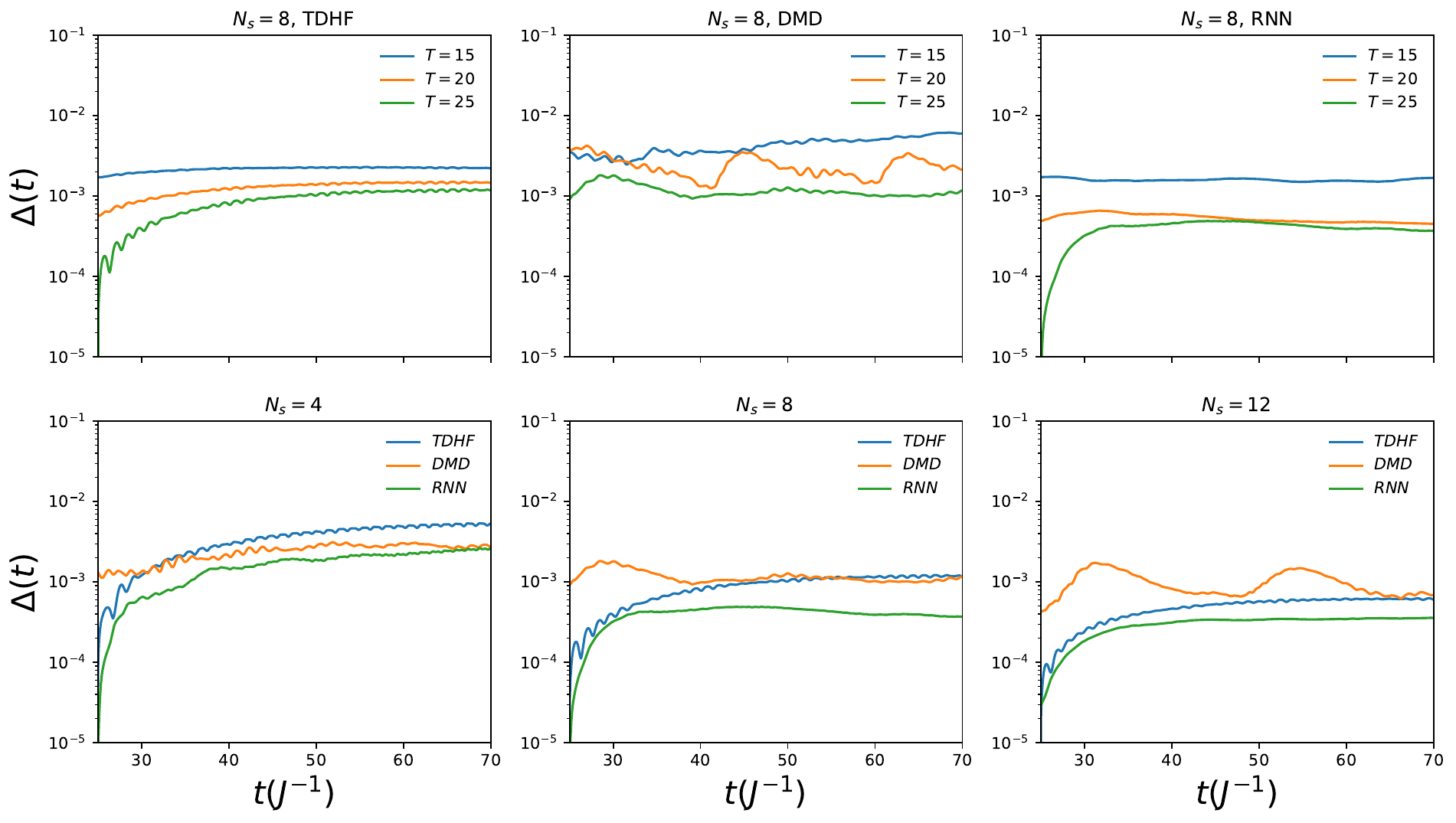}
\end{subfigure}
\captionsetup{justification=justified}
\caption{Error plot of three dynamics extrapolation methods for half-filled Hubbard model with $N_s=4,8,12$, where the learning data length is $T=15,20,25$. The plots in the first row is for the 8-site Hubbard model. They show the numerical convergence of these three methods as we gradually increase the data length. The plots in the second row compare the numerical accuracy of these three methods for the Hubbard model with sites $N_s=4,8,12$, where the training data length is fixed to be $T=25$. One can see that the RNN produces the most accurate prediction for all test models.}
\label{fig:convergence_test_error}
\end{figure}

The first step of our approach is to show whether the dynamics extrapolation methods exhibit numerical convergence. Specifically, we gradually increase the training data length and see whether the predicted Green's function dynamics converge. We performed such numerical convergence tests for the TDHF, DMD, and RNN approaches for Hubbard model with different sites. From Figure \ref{fig:convergence_test_N4}-\ref{fig:convergence_test_N12}, we see that both the RNN and DMD methods show clear numerical convergence to the KBE solution as we gradually increase the learning data length, while the DMD convergence speed is lower than the RNN. A more quantitative assessment for the numerical convergence and accuracy of the dynamics extrapolation result is by evaluating the extrapolation error $\Delta(t)$ defined by: 
\begin{align*}
 \Delta(t)=\displaystyle\frac{1}{N_s^2}\|\rho(t)-\tilde\rho(t)\|_{2,2}=\frac{1}{N_s^2}\left(\sum_{i=1}^{N_s}\sum_{j=1}^{N_s}|\rho_{ij}(t)-\tilde\rho_{ij}(t)|^2\right)^{1/2},
\end{align*}
where $\rho(t)=\rho_{KBE}(t)$ and $\tilde \rho(t)= \rho_{TDHF}(t),\rho_{DMD}(t),\rho_{RNN}(t)$. From Figure \ref{fig:convergence_test_error}, we see more clearly the numerical convergence of the RNN method, and its accuracy over the TDHF and DMD approach. 

The numerical convergence of the RNN approach enables a simple way to predetermine the learning data length for dynamics extrapolation. Namely, if the relative difference between the predicted Green's function generated by RNN with different data lengths, say $T=20$ and $T=25$, is smaller than a certain threshold, we may conclude that the learning data $T=20$ is enough for producing reasonable dynamics predictions. A clear illustrative example is shown in the rightmost subplot of the first row of Figure \ref{fig:convergence_test_error}, from which, we see that the RNN with training data $T=20$ and $T=25$ generates extrapolation results with similar accuracy. Combining with Figure \ref{fig:convergence_test_N8}, we see that with the training data up to $T=20,25$, the RNN indeed makes similar predictions of the Green's function dynamics, and both of them well approximate the ground-truth KBE solution.   

\bibliographystyle{plain}